\documentclass[10pt]{article}
\usepackage{amsmath}
\usepackage[shortlabels]{enumitem}
\usepackage{amsfonts}       
\usepackage{nicefrac}       
\usepackage{bbm}
\usepackage{float}

\usepackage[strict]{changepage}
\usepackage[marginal]{footmisc}
\usepackage{theorem}
\usepackage{mathrsfs}
\usepackage{authblk}
\usepackage{times}
\usepackage{thmtools}
\usepackage{amssymb,array,graphicx,mathtools,tcolorbox,booktabs}
\usepackage[utf8]{inputenc}
\usepackage[T1]{fontenc}

\usepackage[caption=false,font=normalsize,labelfont=sf,textfont=sf]{subfig}
\usepackage{textcomp}
\usepackage{stfloats}
\usepackage{url}
\usepackage{cite}
\usepackage[paperwidth=199.8mm,paperheight=297mm,centering,hmargin=20mm,vmargin=2cm]{geometry}
\usepackage{hyperref}
\hypersetup{colorlinks=true,citecolor=blue,linkcolor=blue,filecolor=blue,urlcolor=blue,breaklinks=true}

\newtheorem{definition}{Definition}
\newtheorem{proposition}{Proposition}

\newtheorem{theorem}[proposition]{Theorem}

\newtheorem{remark}{Remark}
\newtheorem{example}{Example}
\newenvironment{proof}{\noindent \textbf{{Proof~} }}{\hfill $\blacksquare$}

\newcommand{\nc}{\newcommand}
\nc{\rnc}{\renewcommand}
\nc{\lbar}[1]{\overline{#1}}
\nc{\bra}[1]{\langle#1|}
\nc{\ket}[1]{|#1\rangle}
\nc{\ketbra}[2]{|#1\rangle\!\langle#2|}
\nc{\braket}[2]{\langle#1|#2\rangle}

\nc{\proj}[1]{| #1\rangle\!\langle #1 |}
\nc{\avg}[1]{\langle#1\rangle}
\nc{\smfrac}[2]{\mbox{$\frac{#1}{#2}$}}
\nc{\tr}{\operatorname{Tr}}
\nc{\ox}{\otimes}
\nc{\dg}{\dagger}
\nc{\dn}{\downarrow}
\nc{\cA}{{\cal A}}
\nc{\cB}{{\cal B}}
\nc{\cC}{{\cal C}}
\nc{\cD}{{\cal D}}
\nc{\cE}{{\cal E}}
\nc{\cF}{{\cal F}}
\nc{\cG}{{\cal G}}
\nc{\cH}{{\cal H}}
\nc{\cI}{{\cal I}}
\nc{\cJ}{{\cal J}}
\nc{\cK}{{\cal K}}
\nc{\cL}{{\cal L}}
\nc{\cM}{{\cal M}}
\nc{\cN}{{\cal N}}
\nc{\cO}{{\cal O}}
\nc{\cP}{{\cal P}}
\nc{\cQ}{{\cal Q}}
\nc{\cR}{{\cal R}}
\nc{\cS}{{\cal S}}
\nc{\cT}{{\cal T}}
\nc{\cU}{{\cal U}}
\nc{\cV}{{\cal V}}
\nc{\cX}{{\cal X}}
\nc{\cY}{{\cal Y}}
\nc{\cZ}{{\cal Z}}
\nc{\cW}{{\cal W}}

\nc{\argmin}{{\operatorname{argmin}}}
\DeclareMathOperator{\ima}{Im}
\nc{\RR}{{{\mathbb R}}}
\nc{\CC}{{{\mathbb C}}}
\nc{\FF}{{{\mathbb F}}}
\nc{\NN}{{{\mathbb N}}}
\nc{\ZZ}{{{\mathbb Z}}}
\nc{\PP}{{{\mathbb P}}}
\nc{\QQ}{{{\mathbb Q}}}
\nc{\UU}{{{\mathbb U}}}
\nc{\EE}{{{\mathbb E}}}
\nc{\id}{{\operatorname{id}}}

\nc{\CHSH}{{\operatorname{CHSH}}}
\nc{\RecBC}{{{\Theta_{BC|A}}}}
\nc{\RecB}{{{\Theta_{B|A}}}}
\nc{\RecBt}{{{\widetilde\Theta_{B|A}}}}
\nc{\CMC}{{{\text{CMC}}}}
\nc{\QMC}{{{\text{QMC}}}}
\nc{\QQMC}{{{\text{QQMC}}}}

\newcommand{\CPTP}{\text{\rm CPTP}}
\newcommand{\HPTP}{\text{\rm HPTP}}
\newcommand*\samethanks[1][\value{footnote}]{\footnotemark[#1]}

\begin{document}

\title{\textbf{Virtual Quantum Markov Chains}}

\author{Yu-Ao Chen\thanks{Yu-Ao Chen and Chengkai Zhu contributed equally to this work.} }
\author{Chengkai Zhu\samethanks[1]}
\author{Keming~He}
\author{Mingrui~Jing}
\author{Xin~Wang
\thanks{
felixxinwang@hkust-gz.edu.cn}}
\affil{\small Thrust of Artificial Intelligence, Information Hub,\par The Hong Kong University of Science and Technology (Guangzhou), Guangdong 511453, China}

\maketitle

\begin{abstract}
Quantum Markov chains generalize classical Markov chains for random variables to the quantum realm and exhibit unique inherent properties, making them an important feature in quantum information theory. In this work, we propose the concept of \textit{virtual quantum Markov chains} (VQMCs), focusing on scenarios where subsystems retain classical information about global systems from measurement statistics. As a generalization of quantum Markov chains, VQMCs characterize states where arbitrary global shadow information can be recovered from subsystems through local quantum operations and measurements. We present an algebraic characterization for virtual quantum Markov chains and show that the virtual quantum recovery is fully determined by the block matrices of a quantum state on its subsystems. Notably, we find a distinction between two classes of tripartite entanglement by showing that the W state is a VQMC while the GHZ state is not. Furthermore, we introduce the virtual non-Markovianity to quantify the non-Markovianity of a given quantum state, which also assesses the optimal sampling overhead for virtually recovering this state. Our findings elucidate distinctions between quantum Markov chains and virtual quantum Markov chains, extending our understanding of quantum recovery to scenarios prioritizing classical information from measurement statistics.
\end{abstract}

\section{Introduction}
Quantum recovery refers to the ability to reverse the effects of a quantum operation on a quantum state, allowing for the retrieval of the original state~\cite{Petz1986,Petz1988,Barnum2002a}. When this quantum operation involves discarding a subsystem (mathematically represented as a partial trace) of a tripartite quantum state $\rho_{ABC}$, the concept of recoverability becomes intertwined with Quantum Markov Chains~\cite{Hayden2004}. 

A tripartite quantum state $\rho_{ABC}$ is called a  Quantum Markov Chain in order $A\leftrightarrow B\leftrightarrow C$ if there exists a recovery channel $\cR_{B \to BC}$ that can perfectly reconstruct the original whole state from the $B$-part only, i.e.,
\begin{equation}\label{eq:Markov}
\rho_{ABC} =  \cR_{B \to BC}(\rho_{AB}). 
\end{equation}
There are two main ways to characterize quantum Markov chains. The entropic characterization of quantum Markov chains states that a tripartite state $\rho_{ABC}$ is a quantum Markov chain if and only if the conditional mutual information $I(A:C|B)_{\rho}$ is zero~\cite{Petz1986}. The Petz recovery map, a specific quantum channel, can perfectly reverse the action of the partial trace operation for such quantum Markov chains. Moreover, the algebraic characterization of quantum Markov chains is based on the decomposition of the second subsystem~\cite{Hayden2004}, i.e., a tripartite state $\rho_{ABC}$ is a quantum Markov chain if and only if system $B$ can be decomposed into a direct sum tensor product
\begin{equation}
    \cH_B = \underset{j}{\bigoplus}  \mathcal{H}_{b^L_j} \ox \cH_{b^R_j}, ~ \text{s.t.}~ \rho_{ABC} = \underset{j}{\bigoplus} q_j \rho_{Ab^L_j} \ox \rho_{b^R_jC},  \label{Markov_state}
\end{equation} 
with states $\rho_{Ab^L_j}$ on $\mathcal{H}_A \otimes \mathcal{H}_{b^L_j}$ and $\rho_{b^R_jC}$ on $ \mathcal{H}_{b^R_j} \otimes \mathcal{H}_C $ and a probability distribution $\{q_j\}$. 

Quantum states that have a small conditional mutual information are shown to be approximate quantum Markov chains as they can be approximately recovered~\cite{Fawzi2014}. In particular, Fawzi and Renner~\cite{Fawzi2014} shows that for any state $\rho_{ABC}$ there exists a recovery channel $\cR_{B \to BC}$ such that
\begin{align} \label{eq_FR}
I(A:C|B)_{\rho} \geq - \log F\big(\rho_{ABC},\cR_{B\to BC}(\rho_{AB})\big) \, ,
\end{align}
where $F(\cdot,\cdot)$ denotes the fidelity between quantum states. Substantial efforts have been made to further understand the approximate quantum Markov chains and the recoverability in quantum information theory (see, e.g.,~\cite{Sutter2018,Ibinson2007,Brandao2014,Berta2015,Wilde2015,Lami2018a,Seshadreesan2015a,Sutter2017,Sutter2016c}). Moreover, reconstructing a quantum state from local marginals has been explored previously in the context of matrix product operators and finitely correlated states~\cite{Baumgratz_2013,Holz_pfel_2018}, with a more detailed algorithm and error analysis recently~\cite{Fanizza2024}.

\begin{figure}[t]
    \centering
    \includegraphics[width=.55\linewidth]{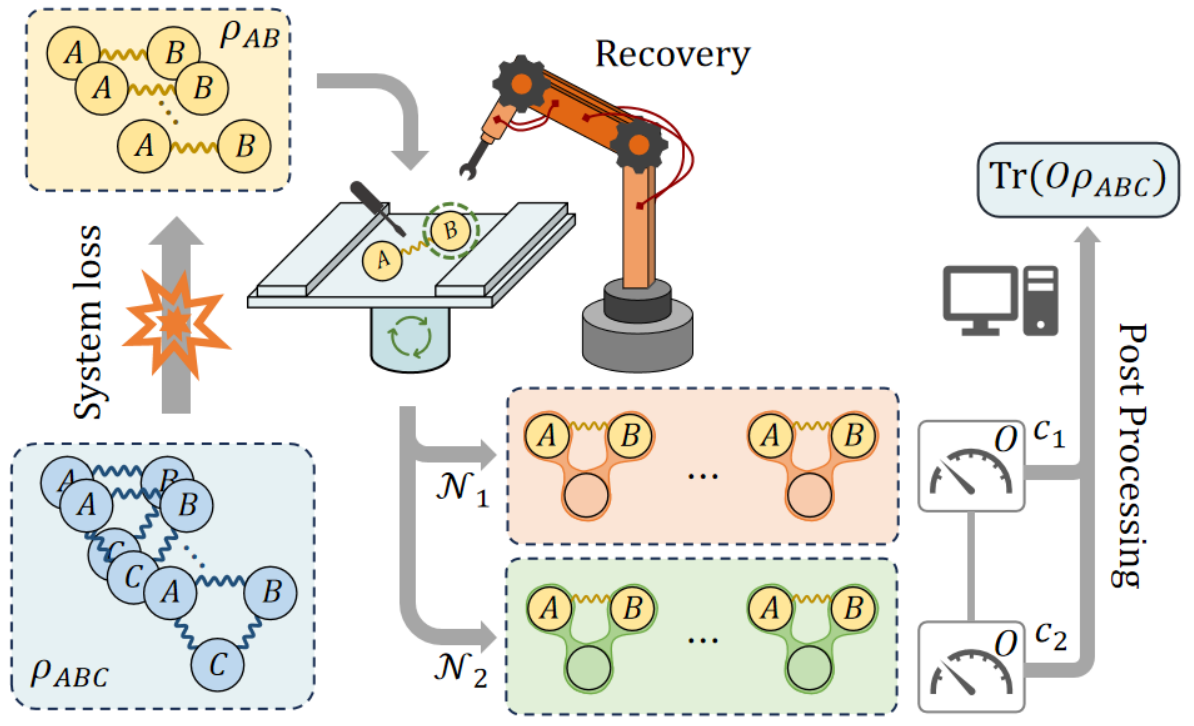}
    \caption{The framework of recovering measurement statistics of a tripartite quantum state with respect to any observable from subsystem $B$ via sampling quantum operations and classical post-processing. The recovery map $\mathscr{R}_{B\rightarrow BC}$ can be simulated via sampling quantum channels $\cN_1$ and $\cN_2$ with probabilities $c_1/(c_1+c_2)$ and $c_2/(c_1+c_2)$, respectively, where $\mathscr{R}_{B\rightarrow BC} = c_1 \cN_1 - c_2 \cN_2$.}
    \label{fig:setting}
\end{figure}

However, can we have a more efficient method for extracting specific information of the original state without fully reconstructing it? Within quantum information processing, the emphasis primarily lies on information gleaned from measurement outcomes, rather than on the quantum state itself. In particular, a quantum state is fundamentally an entity encoding the expectation values for all conceivable observables.
Given this interpretation, it is intuitively expected that reconstructing a quantum state from a local subsystem to a global system should ideally restore the critical information necessary to determine the expectation values for any observables.
For the extraction of classical information from quantum systems, Scott Aaronson proposed the notion of \textit{shadow tomography}~\cite{Aaronson2018b}, which has become an indispensable tool in both quantum computing and quantum information theory~\cite{Huang_2020}. These expectation values, also known as shadow information~\cite{Zhao2022}, have garnered significant interest across various domains, including quantum error mitigation~\cite{Temme2017,Jiang2020,Piveteau2021,Zhao2023}, distributed quantum computing~\cite{Mitarai2021a,Piveteau2022,Yuan2023}, correlation functions~\cite{Buscemi2013}, entanglement detection~\cite{Elben2020,Wang2020,Regula2021a}, quantum broadcasting~\cite{Yao2023,Parzygnat_2024}, and fault-tolerant quantum computing~\cite{Piveteau2021a}. In this context, prior investigations into quantum Markov chains have incompletely elucidated the information recoverability from correlated subsystems. A substantial gap remains in our understanding of how to optimally exploit these local correlations to precisely obtain expectation values for global unknown observables.

In this paper, we introduce the \textit{virtual quantum Markov chains} (VQMCs), which characterize the quantum state whose global shadow information can be recovered from subsystems via local quantum operations and post-processing.
To be specific, a tripartite quantum state $\rho_{ABC}$ is called a virtual quantum Markov chain in order $A\leftrightarrow B\leftrightarrow C$ if there exists a Hermitian preserving map $\mathscr{R}_{B\rightarrow BC}$ such that 
\begin{equation}
    \mathscr{R}_{B\rightarrow BC} (\rho_{AB}) = \rho_{ABC}.
\end{equation}
The above existence of $\mathscr{R}_{B\rightarrow BC}$ ensures that the value of $\tr (O\rho_{ABC})$ can be retrieved for any observable $O$ from $\rho_{AB}$ via statistically simulating $\mathscr{R}_{B\rightarrow BC}$ using quasiprobability decomposition~\cite{Jiang2020} or
measurement-controlled post-processing~\cite{Zhao2023}. The process of a virtual quantum recovery is illustrated in Fig.~\ref{fig:setting}. A fundamental distinction between VQMCs and quantum state reconstruction from local marginals~\cite{Baumgratz_2013} or matrix reconstruction~\cite{Holz_pfel_2018} is that VQMC focuses on recovering the expectation values of a state via sampling and implementing physical operations, while traditional reconstruction methods may apply general linear maps.

We present an algebraic characterization of the VQMCs, offering an easily verifiable criterion for determining a quantum state's qualification as a VQMC. This criterion suggests that a tripartite quantum state can undergo virtual recovery if and only if the kernel of its block matrix on subsystem $B$, conditional on $A$, is included in the kernel of its block matrix on subsystem $BC$, conditional on $A$, (cf.~Theorem~\ref{thm:main_necc_suff}). As a notable application, we show that a W state is a VQMC while a GHZ state is not. 

Furthermore, we propose a protocol for recovering shadow information for arbitrary observables from a VQMC via sampling local quantum operations and classical post-processing. We explore the sampling overhead of the recovery protocol through semidefinite programming (cf.~Sec.~\ref{sec:VNonMark}), shedding light on the distinctions between a quantum Markov chain and a virtual quantum Markov chain. We demonstrate that the optimal sampling overhead for a virtual recovery protocol is additive with respect to the tensor product of states, indicating that a parallel recovering strategy has no advantage over a local protocol, i.e., recovering each state individually. Third, we introduce the approximate virtual quantum Markov chain (cf.~Sec.~\ref{sec:approx vqmc}), in which we can recover the shadow information with respect to any observable approximately. We also characterize the approximate recoverability via semidefinite programming.

\section{Virtual quantum Markov chains}\label{sec:VQMC}

We label different quantum systems by capital Latin letters, e.g., $A, B$. The respective Hilbert spaces for these quantum systems are denoted as $\cH_{A}$, $\cH_{B}$, each with dimension $d_A, d_B$. The set of all linear operators on $\cH_A$ is denoted by $\cL(A)$, with $I_A$ representing the identity operator. We denote by $\cL^{\dag}(A)$ the set of all Hermitian operators on $\cH_A$. In particular, we denote by $\cD(A)\subseteq \cL^{\dag}(A)$ the set of all density operators that are positive semidefinite and trace-one acting on $\cH_A$. Throughout the paper, for a tripartite quantum state $\rho_{ABC} \in \cD(A\ox B\ox C)$, we denote $\rho_{AB} = \tr_C \rho_{ABC}$. 
A linear map transforming linear operators in system $A$ to those in system $B$ is termed a quantum channel if it is completely positive and trace-preserving (CPTP), denoted as $\cN_{A\rightarrow B}: \cL(A)\rightarrow \cL(B)$. The set of all quantum channels from $A$ to $B$ is denoted as $\CPTP(A,B)$.
If a map transforms linear operators in $\cL^{\dag}(A)$ to those in $\cL^{\dag}(B)$ and trace-preserving, it is termed a Hermitian-preserving and trace-preserving (HPTP) map. The set of all HPTP maps from $A$ to $B$ is denoted as $\HPTP(A,B)$.

We start with the definition of a \textit{virtual quantum Markov chain} (VQMC).

\begin{definition}[Virtual quantum Markov chain]
    A tripartite quantum state $\rho_{ABC} \in \cD(A\ox B\ox C)$ is called a virtual quantum Markov chain in order $A\leftrightarrow B\leftrightarrow C$ if there exists a recovery map $\mathscr{R}_{B\rightarrow BC}=\sum_{i} \eta_i \cN^{(i)}_{B\rightarrow BC}$ where $\cN^{(i)}_{B\rightarrow BC} \in \CPTP(B,B\ox C),\eta_i\in \mathbb{R}$ such that
    \begin{equation}
        \mathscr{R}_{B\rightarrow BC}\circ \tr_C(\rho_{ABC}) = \rho_{ABC}.
    \end{equation}
\end{definition}

Equivalently, a state $\rho_{ABC}$ is a VQMC if and only if there exist two quantum channels $\cN_1$ and $\cN_2$ with non-negative real coefficients $c_0,c_1$, such that $\mathscr{R}_{B\to BC} = c_1\cN_1 - c_2\cN_2$ and $\mathscr{R}_{B\rightarrow BC}\circ \tr_C(\rho_{ABC}) = \rho_{ABC}$. By definition, a quantum Markov chain is also trivially included as a virtual quantum Markov chain. Recall that a quantum Markov chain is a state in which the $C$-part can be reconstructed by locally acting on the $B$-part, thereby recovering the complete information of a state. Extending this, we will see that a virtual quantum Markov chain is a state in which the measurement statistics of any observable associated with quantum systems can be reconstructed by exclusively operating on the $B$-part, even when the $C$-part is dismissed. The map $\mathscr{R}_{B\rightarrow BC}$ for a virtual quantum Markov chain is termed a \textit{virtual recovery map}. We note that it is already known that HPTP maps can be written as a weighted difference of CPTP maps and simulated through quasiprobability decomposition (QPD)~\cite{Temme2017,Jiang2020,Piveteau2021,Zhao2022} and measurement-controlled post-processing~\cite{Zhao2023}, so that it is enough to ask for HPTP in the definition of virtual Markov chain. In detail, by sampling quantum channels $\cN_{B\rightarrow BC}^{(i)}$ and performing classical post-processing, we are able to estimate the expectation value of any possible observable in $\rho_{ABC}$. Within this framework, we can accurately retrieve the value of $\tr(O\rho_{ABC})$ for any observable $O$ under a desired error threshold $\epsilon$ by locally operating on the $B$-part of $\rho_{AB}$.

Following the interpretation of a virtual quantum Markov chain, it is natural to expect that a wider class of states can qualify as VQMCs. But does this extended range include all quantum states? We address this question negatively by presenting a necessary and sufficient condition for a state to be a virtual quantum Markov chain. To characterize the structure inherent in a virtual quantum Markov chain, we first introduce the block matrix of a quantum state on subsystems as follows. For a given $d_A\ox d_B$ bipartite quantum state $\rho_{AB}$, we denote $Q_{B}^{(ij)} := \bra{i}_A \rho_{AB} \ket{j}_A$ as its block matrix on subsystem $B$ where $\{\ket{i}_A\}_i$ is the computational basis on subsystem $A$. Consequently, we define the \textit{block matrix of $\rho_{AB}$ on subsystem $B$} as 
\begin{equation}
    \RecB := \left[Q_{B}^{(00)}, Q_{B}^{(01)}, \cdots, Q_{B}^{(d_A-1d_A-1)}\right].
\end{equation}
The kernel, or the null space of $\RecB$, is given by
\begin{equation}
    \ker\RecB := \left\{\mathbf{c} \in \mathbb{C}^{d_A^2 \times 1} \,|\, \RecB \cdot \mathbf{c} = 0\right\},
\end{equation}
and the image of $\RecB$ is given by 
\begin{equation}
    \ima\RecB := \left\{\RecB \cdot \mathbf{c} \, | \, \mathbf{c} \in \mathbb{C}^{d_A^2 \times 1} \right\}.
\end{equation}
Note that it is straightforward to generalize the block matrix on a subsystem for a multipartite quantum state. A tripartite quantum state $\rho_{ABC}$ can simply be written as $\rho_{ABC} = \sum_{ij}\ketbra{i}{j}\ox Q_{BC}^{(ij)}$ and has
\begin{equation}
    \RecBC := \left[Q_{BC}^{(00)}, Q_{BC}^{(01)}, \cdots, Q_{BC}^{(d_A-1d_A-1)}\right].
\end{equation}
Based on the above, the algebraic structure of a VQMC can be characterized as the following theorem.

\begin{theorem}\label{thm:main_necc_suff}
A tripartite quantum state $\rho_{ABC}$ is a virtual quantum Markov chain in order $A\leftrightarrow B\leftrightarrow C$ if and only if 
\begin{equation}
    \ker\RecB \subseteq \ker\RecBC,
\end{equation}
where $\RecB$ and $\RecBC$ are the block matrices of $\rho_{ABC}$ on subsystem $B$ and $BC$, respectively.
\end{theorem}

\begin{proof}
For the "only if" part: By definition, there exists a recovery map $\mathscr{R}_{B\rightarrow BC}$ such that $\mathscr{R}_{B\rightarrow BC}\circ \tr_C(\rho_{ABC}) =\rho_{ABC}$, i.e.
\begin{equation}
    \forall\,i,j,\ \mathscr{R}_{B\rightarrow BC}\left(Q_B^{(ij)}\right) =  Q_{BC}^{(ij)}.
\end{equation}
For any $\mathbf{c}\in\ker\RecB$, it is easy to check that
\begin{equation}
    \RecBC\cdot\mathbf c=\mathscr{R}_{B\rightarrow BC}\left(\RecB\cdot\mathbf c\right)=\mathscr{R}_{B\rightarrow BC}\left(0\right)=0.
\end{equation}
As a result, $\ker\RecB \subseteq \ker\RecBC$ is proved.

For the "if" part: Given a $d_A\ox d_B\ox d_C$ tripartite quantum state $\rho_{ABC}$, regarding $\RecB$ and $\RecBC$ as linear maps from $\mathbb{C}^{d_A^2 \times 1}$ to $\mathbb C^{d_B\times d_B}$ and $\mathbb C^{d_Bd_C\times d_Bd_C}$ respectively, we could check $\tr_C\circ \RecBC=\RecB$, which implies that $\ker\RecBC\subseteq\ker\RecB$. Notice that $\ker\RecB \subseteq \ker\RecBC$ as supposed, we find $\ker\RecB=\ker\RecBC$ and furthermore the linear space $\ima\RecB$ is isomorphic to $\ima\RecBC$. As a result, the surjection $\tr_C$ is just a bijection from $\ima\RecBC$ to $\ima\RecB$. Thus, a linear map $\cM:\ima\RecB\rightarrow\ima\RecBC$ is denoted as the inverse map of $\tr_C:\ima\RecBC\rightarrow\ima\RecB$.
Denote the orthogonal complement space of the image space $\ima\RecB\subseteq \mathbb C^{d_B\times d_B}$ as $\left(\ima\RecB\right)^\perp\subseteq \mathbb C^{d_B\times d_B}$, and define linear maps
\begin{align}
&\mathcal N: \left(\ima\RecB\right)^\perp\rightarrow \mathbb C^{d_Bd_C\times d_Bd_C},~M\mapsto\frac{\tr M}{d_Bd_C} I_{d_Bd_C},\label{eq:def_cN}\\
&\mathscr{R}=\cM\oplus\mathcal N: \mathbb C^{d_B\times d_B}\rightarrow \mathbb C^{d_Bd_C\times d_Bd_C},~\sigma\mapsto\cM\circ\Pi_{\ima\RecB}(\sigma)+\mathcal N\circ\Pi_{\left(\ima\RecB\right)^\perp}(\sigma),
\end{align}
where $\Pi_{\ima\RecB}$ and $\Pi_{\left(\ima\RecB\right)^\perp}$ denote the projection operators for subspaces $\ima\RecB$ and $\left(\ima\RecB\right)^\perp$, respectively. To prove $\rho_{ABC}$ is a virtual quantum Markov chain, we only need to check that
\begin{equation}\label{Eq:vir_rec_ability}
    \mathscr{R}_{B\rightarrow BC}\circ \tr_C(\rho_{ABC}) = \rho_{ABC},
\end{equation}
and $\mathscr{R}_{B\rightarrow BC}$ is both Hermitian-preserving and trace-preserving. For Eq.~\eqref{Eq:vir_rec_ability}, we have
\begin{equation}
\begin{aligned}
\mathscr{R}_{B\rightarrow BC}\circ \tr_C(\rho_{ABC}) = &\sum_{ij}\ketbra{i}{j}_A\ox\mathscr{R}_{B\rightarrow BC}\circ \tr_CQ_{BC}^{(ij)}\\
=&\sum_{ij}\ketbra{i}{j}\ox\Big(\cM\circ\Pi_{\ima\RecB}\circ \tr_CQ_{BC}^{(ij)} +\cN\circ\Pi_{\left(\ima\RecB\right)^\perp}\circ \tr_CQ_{BC}^{(ij)}\Big)\\ 
=&\sum_{ij}\ketbra{i}{j}\ox\left(\cM\circ\tr_CQ_{BC}^{(ij)}+0\right)\\
=&\sum_{ij}\ketbra{i}{j}\ox Q_{BC}^{(ij)}\\
=&\rho_{ABC}.
\end{aligned}
\end{equation}
For `trace-preserving', we can check that for any $\sigma\in \mathbb C^{d_B\times d_B}$. It follows
\begin{equation}
\begin{aligned}
    \tr\circ \mathscr{R}(\sigma)=&\tr\circ\cM\circ\Pi_{\ima\RecB}(\sigma)+\tr\circ \cN \circ\Pi_{\left(\ima\RecB\right)^\perp}(\sigma)\\
    =&\tr_B\circ\left(\tr_C\circ\cM\circ\Pi_{\ima\RecB}(\sigma)\right)  +\tr\left(\frac{\tr\Pi_{(\ima\RecB)^\perp}(\sigma)}{d_Bd_C}I_{BC}\right)\\
    =&\tr_B\circ\left(\Pi_{\ima\RecB}(\sigma)\right)+\tr\circ\Pi_{\left(\ima\RecB\right)^\perp}(\sigma)\\    
    =&\tr\sigma,
\end{aligned}
\end{equation}
where we use the fact that $\forall\,\sigma,\ \Pi_{\ima\RecB}(\sigma)+\Pi_{\left(\ima\RecB\right)^\perp}(\sigma)=\sigma$.

For `Hermitian-preserving', we note that for any $\sigma\in \cL^{\dag}(B)$, we only need to prove both $\cM\circ\Pi_{\ima\RecB}(\sigma)$ and $\cN\circ\Pi_{\left(\ima\RecB\right)^\perp}(\sigma)$ are always Hermitian. Denote 
\begin{equation}
    \Pi_{\ima\RecB}(\sigma)=\sum_{ij}a_{ij}Q_B^{(ij)}.
\end{equation}
Since $Q_B^{(ij)\dagger} = Q_B^{(ji)}$, we have $\left(\ima\RecB\right)^\dagger=\ima\RecB$ and $\left(\ima\RecB\right)^{\perp\dagger}=\left(\ima\RecB\right)^{\perp}$. Moreover, by
\begin{align}
    \Pi_{\ima\RecB}(\sigma)+\Pi_{\left(\ima\RecB\right)^\perp}(\sigma)
    =&\; \sigma = \sigma^\dagger\\
    =&\; \Pi_{\ima\RecB}(\sigma)^\dagger+\Pi_{\left(\ima\RecB\right)^\perp}(\sigma)^\dagger\\
    =&\; \Pi_{\ima\RecB}\left(\Pi_{\ima\RecB}(\sigma)^\dagger\right) + \Pi_{\left(\ima\RecB\right)^\perp}\Big(\Pi_{\left(\ima\RecB\right)^\perp}(\sigma)^\dagger\Big),
\end{align}
we find both $\Pi_{\ima\RecB}(\sigma)$ and $\Pi_{\left(\ima\RecB\right)^\perp}(\sigma)\in\cL^\dag(B)$. Since 
\begin{equation}
    \ima\RecB=\operatorname{Span}_{\mathbb R}\Big(\Big\{Q_B^{(ij)}+Q_B^{(ji)},\sqrt{-1}(Q_B^{(ij)}-Q_B^{(ji)})\Big\}_{ij}\Big),
\end{equation}
we could assume each $a_{ij}^*=a_{ji}$. Thus $\cM\circ\Pi_{\ima\RecB}(\sigma)$ is Hermitian because
\begin{align}
    \left(\cM\circ\Pi_{\ima\RecB}(\sigma)\right)^\dagger
    =&\left(\sum_{ij}a_{ij}\cM \left(Q_B^{(ij)}\right)\right)^\dagger
    =\left(\sum_{ij}a_{ij} Q_{BC}^{(ij)}\right)^\dagger
    =\sum_{ij}a_{ij}^*Q_{BC}^{(ij)\dagger}\\
    =&\sum_{ij}a_{ji}Q_{BC}^{(ji)}
    =\sum_{ij}a_{ji}\cM\left(Q_{B}^{(ji)}\right)
    =\cM\circ\Pi_{\ima\RecB}(\sigma);
\end{align}
$\cN\circ\Pi_{\ima\RecB}(\sigma)$ is Hermitian because $\Pi_{\ima\RecB}(\sigma)$ is Hermitian and $\cN$ is Hermitian-preserving by its definition in Eq.~\eqref{eq:def_cN}. Finally,
we completed the proof for Hermitian-preserving.
\end{proof}

\begin{remark}
Moreover for Theorem~\ref{thm:main_necc_suff}, for a given tripartite quantum state $\rho_{ABC}$, the following are equivalent:
\begin{itemize}
    \item $\rho_{ABC}$ is a virtual quantum Markov chain in order $A\leftrightarrow B\leftrightarrow C$;
    \item $\ker\RecB \subseteq \ker\RecBC$; 
    \item $\ker\RecB=\ker\RecBC$;
    \item there exists a linear map $\cM$ satisfying $\forall\, i,j,\ \mathcal M\left(Q_B^{(ij)}\right)=Q_{BC}^{(ij)}$;
    \item there exists an HPTP map $\cM$ satisfying $\forall\, i,j,\ \mathcal M\left(Q_B^{(ij)}\right)=Q_{BC}^{(ij)}$.
\end{itemize}
\end{remark}

\begin{remark}
We remark that in Ref.~\cite[Theorem 8]{Holz_pfel_2018}, the authors established a necessary and sufficient condition for the existence of local recovery linear maps. Specifically, given a bipartite state $\rho_{XY}$ and its image $\tau_{X'Y'} = (\cN_{X\rightarrow X'}\ox \cN'_{Y\rightarrow Y'})(\rho_{XY})$ under local linear maps $\cN_{X\rightarrow X'}$ and $\cN'_{Y\rightarrow Y'}$, there exist local linear maps $\cR_{X'\rightarrow X},\cR_{Y'\rightarrow Y}$ such that $\rho_{XY} = (\cR_{X'\rightarrow X}\ox \cR_{Y'\rightarrow Y})(\tau_{X'Y'})$ if and only if $\mathrm{OSR}(X:Y)_{\rho} = \mathrm{OSR}(X':Y')_{\tau}$ where the operator Schmidt rank of a linear operator $\rho\in\cL(A\ox B)$ is defined as
\begin{equation}
\mathrm{OSR}(A:B)_{\rho} = \min\Big\{ r ~\Big|~\rho = \sum_{k=1}^{r} G_k' \ox G^{''}_k, G_k'\in \cL(A),G^{''}_k\in \cL(B) \Big\}.
\end{equation}
Compared with this result, what we consider for a VQMC is actually setting $X=A, Y=BC$ with a local map $\cN'_{BC\rightarrow B}(\cdot) = \tr_C(\cdot)$. Denoting $\rho=\sum_{jklr}\rho_{jklr} \ketbra{j}{k}_A\ox 
 \ketbra{l}{r}_B$, we have
\begin{equation}
\Theta_{B|A}=\sum_{jklr}\rho_{jklr} \langle jk|_A\ox \ketbra{l}{r}_B,
\end{equation}
and 
\begin{equation}
    \mathrm{OSR}(A:B)_{\rho_{AB}}=d_A^2-\dim\ker\RecB.
\end{equation}
We could see that
\begin{equation}\label{Eq:OSR_equiv}
    \mathrm{OSR}(A:B)_{\rho_{AB}}=\mathrm{OSR}(A:B\ox C)_{\rho_{ABC}},
\end{equation}
is a special case of the condition in~\cite[Theorem 8]{Holz_pfel_2018}. Notably, Eq.~\eqref{Eq:OSR_equiv} is equivalent to $\ker\RecB=\ker\RecBC$ provided in Theorem~\ref{thm:main_necc_suff} as $\ker\RecBC\subseteq\ker\RecB$ always holds. Both conditions characterize when a state $\rho_{ABC}$ is a virtual quantum Markov chain. A similar criterion was also established for the reconstruction of quantum states that are well approximated by matrix product operators~\cite{Baumgratz_2013}. However, our contribution here is to further extend the result in~\cite[Theorem 8]{Holz_pfel_2018} to the existence of an HPTP recovery map for the partial trace operation. 
This HPTP property, inspired by probabilistic error cancellation techniques in quantum error mitigation~\cite{Temme2017}, is physically more relevant than general linear maps and leads to several important properties that we explore in subsequent sections.
\end{remark}

Theorem~\ref{thm:main_necc_suff} provides an easy-to-check criterion for a state being a VQMC. A straightforward sufficient condition arises: if $\big\{Q_{B}^{(ij)}\big\}_{ij}$ for a state $\rho_{ABC}$ is linear independent, then $\rho_{ABC}$ is a VQMC. Utilizing this theorem, we reveal that the collection of virtual quantum Markov chains does not constitute a convex set.

\begin{proposition}\label{prop:nonconvex}
    The set of all virtual quantum Markov chains is non-convex.
\end{proposition}

\begin{proof}
Consider the following two tripartite quantum states
\begin{equation}
    \rho = \frac{1}{2} \big(\ketbra{\psi_0}{\psi_0} + \ketbra{\psi_1}{\psi_1}\big),
\end{equation}
where $\ket{\psi_0} = \ket{000},~\ket{\psi_1} = \ket{101}$. It is easy to see that $\ketbra{\psi_0}{\psi_0}$ and $\ketbra{\psi_1}{\psi_1}$ are quantum Markov chains as they are all product states. However, for $\rho$, we have
\begin{equation}
    Q_{BC}^{(00)} = \ketbra{00}{00},~Q_{BC}^{(11)} = \ketbra{01}{01}
\end{equation}
and
\begin{equation}
    Q_B^{(00)} = Q_B^{(11)} = \ketbra{0}{0}.
\end{equation}
Therefore, $\ker\RecBC = \mathbf{0}$, but there exists $\mathbf{c} \neq \mathbf{0}$ satisfying $\RecB \cdot \mathbf{c} = 0$. By Theorem~\ref{thm:main_necc_suff}, $\rho$ is not a virtual quantum Markov chain, which implies that the set of all VQMCs is non-convex. 
\end{proof}

The non-convex structure of virtual quantum Markov chains aligns with that of quantum Markov chains, suggesting the persistence of non-convex behavior in Markovian dynamics even when we are only concerned with measurement statistics of quantum states. 

\begin{remark}
Recall that the classicality of a quantum state does not imply that it is a Markov chain. We remark that this holds true for virtual quantum Markov chains as well. That is, even if a state is classical in each subsystem, as in the case of $\rho = \frac{1}{2}(\ketbra{000}{000} + \ketbra{101}{101})$ in the proof of Proposition~\ref{prop:nonconvex}, it may not necessarily constitute a virtual quantum Markov chain.
\end{remark}

To deepen our understanding of virtual quantum Markov chains, we explore the structure of essential tripartite states. The W state and the GHZ state are two representative non-biseparable states that cannot be transformed (not even probabilistically) into each other by local quantum operations~\cite{D_r_2000}. They play important roles in various quantum information tasks, including quantum communication~\cite{zukowski1998quest,qin2017dynamic}, quantum key distribution~\cite{hwang2011quantum}, and quantum algorithms~\cite{roos2004control}. Remarkably, neither the W state nor the GHZ state is a quantum Markov chain. However, in the realm of virtual quantum Markov chains, a divergence appears: the W state aligns with the characteristics of a virtual quantum Markov chain, whereas the GHZ state does not. This distinction elucidates the complex nature of these states within this broader context of quantum Markovian dynamics.

\paragraph{W state}
A generalized W state is a three-qubit entangled quantum state defined by
\begin{equation}\label{Eq:Wstate}
    \ket{W_{\alpha_0,\alpha_1}} = \sqrt{\alpha_0}\ket{001} + \sqrt{\alpha_1}\ket{010} + \sqrt{1-\alpha_0 - \alpha_1}\ket{100},
\end{equation}
where $\alpha_1\neq 0$ and $1-\alpha_0-\alpha_1\neq 0$. For $\rho_{ABC} = \ketbra{W_{\alpha_0,\alpha_1}}{W_{\alpha_0,\alpha_1}}$, we can calculate
\begin{align}
    &Q_B^{(00)} =\left(\begin{array}{cc}
         \alpha _0 &  0 \\
         0 & \alpha _1 \\
    \end{array}\right),
    Q_B^{(01)}=\sqrt{1-\alpha _0-\alpha _1}\left(\begin{array}{cc}
         0 & 0 \\
         \sqrt{\alpha _1} & 0\\
    \end{array}\right),\\
    &Q_B^{(10)}=\sqrt{1-\alpha _0-\alpha _1}\left(\begin{array}{cc}
         0 & \sqrt{\alpha _1}\\
         0 & 0 \\
    \end{array}\right),\\
    &Q_B^{(11)}=\left(\begin{array}{cc}
         1-\alpha _0-\alpha _1 & 0 \\
         0 & 0  \\
    \end{array}\right).
\end{align}
Notice that matrices in $\{Q_{B}^{(ij)}\}_{ij}$ are linear independent and form a basis for $\mathbb{C}^{2\times 2}$. It follows that $\ker\RecB = \ker\RecBC = \mathbf{0}$. 
Therefore, a three-qubit generalized W state is a virtual quantum Markov chain. Particularly, the virtual recovery map $\mathscr{R}_{B\rightarrow BC}$ has a Choi representation as
\begin{equation}
J_{\mathscr{R}} = \frac{\ketbra{0}{0}\otimes Q^{(11)}_{BC}}{1-\alpha_0-\alpha_1}+
\frac{\ketbra{0}{1}\otimes Q^{(10)}_{BC}}{\sqrt{\alpha_1(1-\alpha_0-\alpha_1)}} + \frac{\ketbra{1}{0}\otimes Q^{(01)}_{BC}}{\sqrt{\alpha_0(1-\alpha_0-\alpha_1)}} + \ketbra{1}{1}\otimes\frac{(1-\alpha_0-\alpha_1)Q^{(00)}_{BC}-\alpha_0Q^{(11)}_{BC}}{\alpha_1(1-\alpha_0-\alpha_1)}.
\end{equation}
where $Q^{(ij)}_{BC}$ is the block matrix of $\ketbra{W_{\alpha_0,\alpha_1}}{W_{\alpha_0,\alpha_1}}$ on subsystem $BC$.

\paragraph{GHZ state}
A three-qubit GHZ state is a tripartite entangled quantum state defined by
\begin{equation}\label{Eq:GHZstate}
    \ket{GHZ} = \frac{1}{\sqrt{2}}(\ket{000}+\ket{111}).
\end{equation}
For $\rho_{ABC} = \ketbra{GHZ}{GHZ}$, we have
\begin{align}
    &Q_{BC}^{(00)}=\ketbra{00}{00}/2,~
    Q_{BC}^{(01)}=\ketbra{00}{11}/2,~Q_{BC}^{(10)}=\ketbra{11}{00}/2,~
    Q_{BC}^{(11)}=\ketbra{11}{11}/2,\\
    &Q_{B}^{(00)} =\ketbra{0}{0}/2,~
    Q_{B}^{(01)}=0,~Q_{B}^{(10)}=0,~
    Q_{B}^{(11)}=\ketbra{1}{1}/2.
\end{align}
It is easy to check that there is a $\mathbf{c}\in \mathbb{C}^{1\times 4}$ such that $\RecB \cdot \mathbf{c} = 0$ but $\RecBC \cdot \mathbf{c} \neq 0$. By Theorem~\ref{thm:main_necc_suff}, the GHZ state is not a virtual quantum Markov chain. It indicates that even when we are only interested in measurement statistics of a GHZ state, we still cannot locally recover the information after discarding the subsystem $C$.

In the above, we demonstrate that a three-qubit W state is a virtual quantum Markov chain, but a three-qubit GHZ state is not. This distinction underscores the essential properties of virtual quantum Markov chains. This suggests that the nature and distribution of quantum entanglement within a system could have profound implications for its Markovian properties when we are only concerned with extracting expectation value, e.g., $\tr(O\rho)$. Note that after taking a partial trace on system $C$ for the W states, the reduced density operator contains a residual EPR entanglement. The robustness of W-type entanglement contrasts strongly with the GHZ state, which is fully separable after the loss of one qubit. These findings highlight the need for a deeper understanding of the interplay between multipartite quantum entanglement and Markovian dynamics. 

Furthermore, we investigate the robustness of these virtual quantum Markov chains regarding specific quantum noise. For example, consider a three-qubit W state and a three-qubit GHZ state affected by three-qubit depolarizing channels $\cD_{p}(\cdot) = (1-p)(\cdot) + p I_8/8$ with a noise rate $p\in[0,1]$. We denote the respective states as
\begin{equation}\label{Eq:depo_W_GHZ}
W(p) = (1-p)\ketbra{W}{W} + p \frac{I_8}{8}, G(p) = (1-p)\ketbra{GHZ}{GHZ} + p \frac{I_8}{8},
\end{equation}
where $\ket{W}$ is defined in Eq.~\eqref{Eq:Wstate} with $\alpha_0 = \alpha_1 = 1/3$ and $\ket{GHZ}$ is defined in Eq.~\eqref{Eq:GHZstate}. Then, we have the following results.

\begin{example}\label{exm:W_I}
Let $\ket{W}$ be a three-qubit W state. $W(p) = (1-p)\ketbra{W}{W} + p I_8/8$ is a virtual quantum Markov chain for $p \in [0, 1]$.
\end{example}

\begin{proof}
For $W(p) = (1-p)\ketbra{W}{W} + p I_8/8$ with $\ket{W} = (\ket{001} + \ket{010} + \ket{100})/\sqrt{3}$, we can calculate its block matrices on subsystem $BC$ and $B$ as
\begin{align}
    &Q_{BC}^{(00)}=\left(\begin{array}{cccc}
         p/8 & 0 & 0 & 0 \\
         0 & (1-p)/3 + p/8 & (1-p)/3 & 0 \\
         0 & (1-p)/3 & (1-p)/3+ p/8 & 0 \\
         0 & 0 & 0 & p/8 \\
    \end{array}\right),\\
    &Q_{BC}^{(01)}=\left(\begin{array}{cccc}
         0 & 0 & 0 & 0 \\
         (1-p)/3 & 0 & 0 & 0 \\
         (1-p)/3 & 0 & 0 & 0 \\
         0 & 0 & 0 & 0 \\
    \end{array}\right),\\
    &Q_{BC}^{(10)}=\left(\begin{array}{cccc}
         0 & (1-p)/3 & (1-p)/3 & 0 \\
         0 & 0 & 0 & 0 \\
         0 & 0 & 0 & 0 \\
         0 & 0 & 0 & 0 \\
    \end{array}\right),\\
    &Q_{BC}^{(11)}=\left(\begin{array}{cccc}
         (1-p)/3 + p/8 & 0 & 0 & 0 \\
         0 & p/8 & 0 & 0 \\
         0 & 0 & p/8 & 0 \\
         0 & 0 & 0 & p/8 \\
    \end{array}\right),\\
    &Q_{B}^{(00)} =\left(\begin{array}{cc}
         (1-p)/3 + p/4 &  0 \\
         0 & (1-p)/3 + p/4\\
    \end{array}\right),~
    Q_{B}^{(01)}=\left(\begin{array}{cc}
         0 & 0 \\
         (1-p)/3 & 0\\
    \end{array}\right),\\
    &Q_{B}^{(10)}=\left(\begin{array}{cc}
         0 & (1-p)/3\\
         0 & 0 \\
    \end{array}\right),~
    Q_{B}^{(11)}=\left(\begin{array}{cc}
         (1-p)/3 + p/4 & 0 \\
         0 & p/4 \\
    \end{array}\right).
\end{align}
We can check that $\{Q_{B}^{(ij)}\}$ is linear independent. By Theorem~\ref{thm:main_necc_suff}, $W(p)$ is a virtual Markov chain.
\end{proof}

\begin{example}\label{exm:GHZ_I}
    Let $\ket{GHZ}$ be a three-qubit GHZ state. $G(p) = (1-p)\ketbra{GHZ}{GHZ} + p I_8/8$ is not a virtual quantum Markov chain for $p \in [0, 1)$.
\end{example}

\begin{proof}
For $G(p) = (1-p)\ketbra{GHZ}{GHZ} + p I_8/8$ with $\ket{GHZ}=(\ket{000}+\ket{111})/\sqrt{2}$, we can calculate its block matrices on subsystem $BC$ and $B$ as
\begin{equation}
\begin{aligned}
    &Q_{BC}^{(00)}=\frac{(1-p)}{2}\ketbra{00}{00} + \frac{p}{8}I_4, Q_{BC}^{(01)}=\frac{(1-p)}{2}\ketbra{00}{11}, \\
    &Q_{BC}^{(10)}=\frac{(1-p)}{2}\ketbra{11}{00}, Q_{BC}^{(11)}=\frac{(1-p)}{2}\ketbra{11}{11} + \frac{p}{8}I_4,\\
    &Q_B^{(00)} =\frac{(1-p)}{2}\ketbra{0}{0}+\frac{p}{4}I_2,
    Q_{B}^{(01)}=0,
    Q_{B}^{(10)}=0,\\
    &Q_{B}^{(11)}=\frac{(1-p)}{2}\ketbra{1}{1}+\frac{p}{4}I_2.
\end{aligned}
\end{equation}
Since $Q_{B}^{(01)}= Q_{B}^{(10)} =0$, it is easy to find a $\mathbf{c}\in \mathbb{C}^{1\times 4}$ such that $\RecB \cdot \mathbf{c} = 0$ but $\RecBC \cdot \mathbf{c} \neq 0$. By Theorem~\ref{thm:main_necc_suff}, $G(p)$ is not a virtual quantum Markov chain unless $p = 1$.
\end{proof}

The above examples reveal intrinsic properties inherent in virtual quantum Markov chains. Notice that the maximally mixed state $I_8/8$ is a quantum Markov chain. Example~\ref{exm:W_I} and Example~\ref{exm:GHZ_I} show that the W state and GHZ state maintain their properties of virtual recoverability against depolarizing noise. Example~\ref{exm:W_I} demonstrates that there are cases where a convex combination of VQMCs is still a VQMC even though it is not generally valid. We note that although a GHZ state cannot turn into a VQMC when it is mixed with a maximally mixed state, as shown in Example~\ref{exm:GHZ_I}, it becomes a VQMC when mixed with a W state, as the following example.

\begin{example}\label{exm:ghz_mix_w}
    Let $\ket{GHZ}$ and $\ket{W}$ be the three-qubit GHZ state and W state. For $p\in[0,1],\ GW(p) = p\ketbra{GHZ}{GHZ} + (1-p)\ketbra{W}{W}$ is not a virtual quantum Markov chain if and only if $p=1$ or $7-3\sqrt{5}$.
\end{example}

\begin{proof}
For $GW(p) = p\ketbra{GHZ}{GHZ} + (1-p)\ketbra{W}{W}$ with $\ket{GHZ}=(\ket{000}+\ket{111})/\sqrt{2}$ and $\ket{W} = (\ket{001} + \ket{010} + \ket{100})/\sqrt{3}$, we can calculate its block matrices on subsystem $BC$ and $B$ as
\begin{align}
    &Q_{BC}^{(00)}=\left(\begin{array}{cccc}
        \frac{p}{2} & 0 & 0 & 0 \\
        0 & \frac{1-p}{3} & \frac{1-p}{3} & 0 \\
        0 & \frac{1-p}{3} & \frac{1-p}{3} & 0 \\
        0 & 0 & 0 & 0 \\
    \end{array}\right),~
    Q_{BC}^{(01)}=\left(\begin{array}{cccc}
        0 & 0 & 0 & \frac{p}{2} \\
        \frac{1-p}{3} & 0 & 0 & 0 \\
        \frac{1-p}{3} & 0 & 0 & 0 \\
        0 & 0 & 0 & 0 \\
    \end{array}\right),\\
    &Q_{BC}^{(10)}=\left(\begin{array}{cccc}
        0 & \frac{1-p}{3} & \frac{1-p}{3} & 0 \\
        0 & 0 & 0 & 0 \\
        0 & 0 & 0 & 0 \\
        \frac{p}{2} & 0 & 0 & 0 \\
    \end{array}\right),~
    Q_{BC}^{(11)}=\left(\begin{array}{cccc}
        \frac{1-p}{3} & 0 & 0 & 0 \\
        0 & 0 & 0 & 0 \\
        0 & 0 & 0 & 0 \\
        0 & 0 & 0 & \frac{p}{2} \\
    \end{array}\right),\\
    &Q_{B}^{(00)} =\left(\begin{array}{cc}
        \frac{2+p}{6} & 0 \\
        0 & \frac{1-p}{3} \\
    \end{array}\right),~
    Q_{B}^{(01)}=\left(\begin{array}{cc}
         0 & 0 \\
         \frac{1-p}{3} & 0\\
    \end{array}\right),\\
    &Q_{B}^{(10)}=\left(\begin{array}{cc}
         0 & \frac{1-p}{3}\\
         0 & 0 \\
    \end{array}\right),~
    Q_{B}^{(11)}=\left(\begin{array}{cc}
        \frac{1-p}{3} & 0 \\
        0 & \frac{p}{2} \\
    \end{array}\right).
\end{align}
Notice that $\ker\RecB\ne0$ if only if $\frac{1-p}{3}=0$ or $\frac{2+p}{6}\frac{p}{2}=\frac{1-p}{3}\frac{1-p}{3}$, i.e. $p=1$ or $7-3\sqrt{5}$. If $p=1$ or $7-3\sqrt{5}$, it follows that $\ker\RecBC = 0$. Thus, we have $\ker\RecB \subseteq \RecBC$ if and only if $p=1$ or $7-3\sqrt{5}$, which completes the proof by Theorem~\ref{thm:main_necc_suff}.
\end{proof}

\paragraph{VQMC and quantum conditional mutual information}
Besides algebraic characterization, it is noteworthy that classical Markov chains and quantum Markov chains are interconnected with entropy measures, specifically, conditional mutual information and quantum conditional mutual information, respectively. However, we remark that a virtual quantum Markov chain no longer maintains an intrinsic connection with the quantum conditional mutual information. Herein, we present two tripartite states with the same quantum conditional mutual information, but one is a VQMC, and the other is not.

\begin{example}
Consider two pure three-qubit quantum states given by
\begin{equation}
\ket{\psi_1} = \frac{1}{\sqrt{3}}(\ket{010} + \ket{101} + \ket{110}),~\ket{\psi_2} = \frac{1}{\sqrt{3}}(\ket{010} + \ket{011} + \ket{100}).
\end{equation}
It is easy to check that $I(A:C|B)_{\psi_1} = I(A:C|B)_{\psi_2} = h_2( \frac{1}{2} + \frac{\sqrt{5}}{6})$ where $h_2(p)$ is the binary entropy. Nonetheless, $\ket{\psi_1}$ \textbf{is not} a virtual quantum Markov chain, and $\ket{\psi_2}$ is a virtual quantum Markov chain.
\end{example}

\begin{proof}
For $\ketbra{\psi_1}{\psi_1}$, we have
\begin{equation}
\begin{aligned}
    &Q^{(00)}_{BC}=\frac{1}{3}\ketbra{10}{10},~Q^{(01)}_{BC}=\frac{1}{3}\ket{10}(\bra{01}+\bra{10}),\\
    &Q^{(10)}_{BC}=\frac{1}{3}(\ket{10}+\ket{01})\bra{10}, \\
    &Q^{(11)}_{BC}=\frac{1}{3}(\ket{10}+\ket{01})(\bra{01}+\bra{10}),
\end{aligned}
\end{equation}
and
\begin{equation}
    Q^{(00)}_{B} = Q^{(01)}_{B} = Q^{(10)}_{B} = \frac{1}{3}\ketbra{1}{1}, Q^{(11)}_{B} =\frac{1}{3}(\ketbra{0}{0} + \ketbra{1}{1}).
\end{equation}
It is direct to check for state $\psi_1$, there is a $\mathbf{c}\in \mathbb{C}^{1\times 4}$ such that $\RecB \cdot \mathbf{c} = 0$ while $\RecBC \cdot \mathbf{c}\neq 0$. Hence, state $\ket{\psi_1}$ is not a virtual quantum Markov chain. On the other hand, for state $\ketbra{\psi_2}{\psi_2}$, we have
\begin{equation}
\begin{aligned}
    &Q^{(00)}_{BC} = \frac{1}{3}(\ket{10}+\ket{11})(\bra{10}+\bra{11}), Q^{(11)}_{BC}=\frac{1}{3}\ketbra{00}{00},\\
    &Q^{(01)}_{BC}=\frac{1}{3}(\ket{10}+\ket{11})\bra{00}, Q^{(10)}_{BC}=\frac{1}{3}\ket{00}(\bra{10}+\bra{11}),
\end{aligned}
\end{equation}
and
\begin{equation}
\begin{aligned}
&Q^{(00)}_B = \frac{2}{3}\ketbra{1}{1}, Q^{(01)}_B = \frac{1}{3}\ketbra{1}{0},\\
&Q^{(10)}_B = \frac{1}{3}\ketbra{0}{1},Q^{(11)}_B = \frac{1}{3}\ketbra{0}{0}.
\end{aligned}
\end{equation}
We find sub-matrices $\{Q^{(ij)}_B\}_{ij}$ are linear independent and form a basis of $\mathbb{C}^{2 \times 2}$. Therefore, $\ker\RecB = \ker\RecBC = \mathbf{0}$ and $\ket{\psi_2}$ is a virtual Markov chain.
\end{proof}

We have seen these two states, despite possessing identical quantum conditional mutual information, exhibit differing characteristics - one constitutes a virtual quantum Markov chain, while the other does not. This discrepancy underscores the divergence between quantum conditional mutual information and the properties of virtual quantum Markov chains, indicating the unique structure of a VQMC different from a quantum Markov chain.

\section{Quantifying Non-Markovianity}\label{sec:VNonMark}

For a state not being a quantum Markov chain, its non-Markovianity has been extensively studied by characterizing how a lost quantum system can be recovered from a correlated subsystem approximately~\cite{Sutter2018,fawzi2015quantum}. Here, we introduce a quantity for characterizing the non-Markovianity of a state considering its shadow information recoverability.

For an arbitrary VQMC, we have a recovery map $\mathscr{R}_{B\rightarrow BC}=\sum_i \eta_i\cN_{B\rightarrow BC}^{(i)}$ where each $\cN_{B\rightarrow BC}^{(i)}$ is a quantum channel. With this decomposition, we can simulate the recovery of the expectation value by the probabilistic sampling method~\cite{Jiang2020}. 
Specifically, in each of the $S$ sampling rounds, a quantum channel $\cN_{B\rightarrow BC}^{(s)}$ is randomly selected from ${\cN_{B\rightarrow BC}^{(i)}}_i$ with probabilities ${|\eta_i|/\gamma}$, where $\gamma=\sum_i|\eta_i|$. Applying $\cN_{B\rightarrow BC}$ to subsystem $B$ of the corrupted state $\rho_{AB}$ and measuring the entire state in the eigenbasis of observable $O$ yields an estimation $\frac{\gamma}{S}\sum_{s=1}^S {\rm sgn}(\eta^{(s)}) \lambda(o^{(s)})$ for the expectation value $\tr(O\rho_{ABC})$. To achieve an estimation within an error $\epsilon$ with a probability no less than $1-\delta$, the number of total sampling times $S$ is estimated by Hoeffding's inequality~\cite{hoeffding2012collected} as $S \geq 2\gamma^2 \log(2/\delta)/\epsilon^2$. We note that this probabilistic sampling approach could enable the virtual recovery to serve as an intermediate step for subsequent quantum operations. For example, we can recover entropy-related quantities of a VQMC (see, e.g.,~\cite{acharya2019measuring,subramanian2021quantum,wang2023quantum,wang2024new,haug2024efficient}), as the HPTP property of a recovery map ensures compatibility with follow-up quantum operations by focusing on the measurement statistics. Remarkably, the total sampling times are governed by $\gamma$, which serves as a non-Markovianity quantifier as it quantitatively characterizes the disparity between a VQMC and a QMC.

\begin{definition}[Virtual non-Markovianity]\label{def:simu_cost}
Given a tripartite state $\rho_{ABC}$ with $\rho_{AB}=\tr_C\rho_{ABC}$, the virtual non-Markovianity is defined as
\begin{equation}
\nu(\rho_{ABC})=\log \min \Big\{c_1+c_2~\big|~(c_1\cN_1-c_2\cN_2)(\rho_{AB}) = \rho_{ABC},~c_{1,2}\geq 0,~ \cN_{1,2}\in\CPTP(B,B\ox C) \Big\}.
\end{equation}
$\nu(\rho_{ABC})$ is set to be infinity if $\rho_{ABC}$ is not a VQMC.
\end{definition}

This quantifier for non-Markovianity bears nice properties. Firstly, it is closely related to the physical implementability of HPTP maps~\cite{Jiang2020} and can be determined via the following semidefinite programmings (SDPs)~\cite{boyd2004convex,Mironowicz_2024}, both of which evaluate to $2^{\nu(\rho_{ABC})}$.

\begin{equation}\label{sdp:pri_dual}
    \begin{aligned}
    &\underline{\textbf{Primal Program}}\\
    \min &\;\; c_1 + c_2\\
     {\rm s.t.}
    &\; J_{1}\geq 0, J_{2}\geq 0, \\
    &\; J_{BB'C} = J_1 - J_2,\\
    &\; \tr_{B'C} J_{1} = c_1 I_B,\tr_{B'C} J_{2} = c_2 I_B,\\
    &\; \tr_B(\rho_{AB}^{T_B}\ox I_{B'C})(I_A\ox J_{BB'C}) = \rho_{AB'C},
    \end{aligned}
    \begin{aligned}
    &\underline{\textbf{Dual Program}}\\
    \max &\,  \tr(K_{AB'C}\rho_{ABC})\\
     {\rm s.t.} 
    &\; \tr M_{B} = 1, \\
    &\; \tr N_{B} = 1,\\
    &\; \tr_A(K_{AB'C} \rho_{AB}^{T_B}) \leq M_B\ox I_{B'C},\\
    &\; \tr_A(K_{AB'C} \rho_{AB}^{T_B}) \geq -N_B\ox I_{B'C},
    \end{aligned}
\end{equation}
where $T_B$ denotes taking partial transpose on the subsystem $B$ as $(\ketbra{i_A j_B}{k_A l_B})^{T_B} = \ketbra{i_A l_B}{k_A j_B}$, and the subsystem $B'$ is isomorphic to $B$. In the primal program, the constraint $\tr_B(\rho_{AB}^{T_B}\ox I_{B'C})(I_A\ox J_{BB'C}) = \rho_{AB'C}$ ensures that $J_{BB'C}$ is the recovery map, and we used the fact that every HPTP map can be decomposed into a linear combination of CPTP maps. The derivation of the dual SDP can be found in Appendix~\ref{appendix:sdp_sampling}. Secondly, it is faithful, i.e., $\nu(\rho_{ABC})\geq 0$ and $\nu(\rho_{ABC}) = 0$ if and only if $\rho_{ABC}$ is a quantum Markov chain. 

The virtual non-Markovianity quantitatively characterizes how far a VQMC state is from being a QMC, considering the optimal sampling overhead of the virtual recovery map. According to the result in~\cite[Theorem 3]{Regula2021a}, it can be equivalently expressed as $\nu(\rho_{ABC}) = \log \min \|\mathscr{R}_{B\rightarrow BC}\|_{\diamond}$ where the minimization ranges over all possible virtual recovery maps of $\rho_{ABC}$. Furthermore, for any virtual recovery map $\mathscr{R}_{B\rightarrow BC}(\rho_{AB}) = \rho_{ABC}$ and any CPTP maps $\cN_{B\rightarrow BC}$, we have
\begin{equation}
    \nu(\rho_{ABC}) \leq \log \left(\big\|\mathscr{R}_{B\rightarrow BC} - \cN_{B\rightarrow BC}\big\|_{\diamond} + \|\cN_{B\rightarrow BC}\|_{\diamond}\right) =\log \left(\big\|\mathscr{R}_{B\rightarrow BC} - \cN_{B\rightarrow BC}\big\|_{\diamond} + 1\right).
\end{equation}
$\nu(\rho_{ABC})$ is thus related to the deviation of a virtual recovery map from CPTP maps in terms of the diamond norm distance. Notice that both the virtual non-Markovianity and the conditional mutual information are zero when the state is a QMC. The relationship between non-Markovianity and conditional mutual information remains an intriguing avenue for further exploration, for which we provide a numerical observation in Appendix~\ref{appendix:VNM_vs_CMI}. Notably, we show that the virtual non-Markovianity is additive with respect to the tensor product of two VQMCs.

\begin{proposition}
    For two tripartite virtual quantum Markov chains $\rho_{ABC}$ and $\sigma_{\hat{A}\hat{B}\hat{C}}$, the virtual non-Markovianity is additive, i.e., $\nu(\rho_{ABC}\ox\sigma_{\hat{A}\hat{B}\hat{C}}) = \nu(\rho_{ABC}) + \nu(\sigma_{\hat{A}\hat{B}\hat{C}})$.
\end{proposition}

\begin{proof}
    We will prove $2^{\nu(\rho_{ABC}\ox\sigma_{\hat{A}\hat{B}\hat{C}})} = 2^{\nu(\rho_{ABC})} \cdot 2^{\nu(\sigma_{\hat{A}\hat{B}\hat{C}})}$ using the primal and dual SDP in Eq.~\eqref{sdp:pri_dual}. For `$\leq$': Assume $\{c_1,c_2,J_1,J_2\}$ and $\{\hat{c}_1,\hat{c}_2,\hat{J_1},\hat{J_2}\}$ are feasible solutions for the primal SDP for $\rho_{ABC}$ and $\sigma_{\hat{A}\hat{B}\hat{C}}$, respectively. Let
    \begin{equation}
    \begin{aligned}
        &\widetilde{c_1} = c_1\hat{c}_1+c_2\hat{c}_2, \quad \widetilde{c_2} =c_1\hat{c}_2+c_2\hat{c}_1,\\
        &\widetilde{J_1} = J_1\ox\hat{J}_1 + J_2\ox\hat{J}_2, \quad \widetilde{J_2} = J_1\ox\hat{J}_2 + J_2\ox\hat{J}_1.
    \end{aligned}
    \end{equation}
    Then we have $J_{B\hat{B}B'\hat{B}'C\hat{C}} = \widetilde{J_1} - \widetilde{J_2} = (J_1-J_2)\ox(\hat{J}_1-\hat{J}_2)$. We can check that 
    \begin{equation}
        \tr_{B\hat{B}} \left[\left((\rho_{AB}\ox\sigma_{\hat{A}\hat{B}})^{T_{B\hat{B}}}\ox I_{B'\hat{B}'C\hat{C}}\right)(I_{A\hat{A}}\ox J_{B\hat{B}B'\hat{B}'C\hat{C}})\right]= \rho_{ABC}\ox\sigma_{\hat{A}\hat{B}\hat{C}}
    \end{equation}
    and $\tr_{B'\hat{B}'C\hat{C}} \widetilde{J_1}= (c_1\hat{c}_1+c_2\hat{c}_2)I_{B\hat{B}},\, \tr_{B'\hat{B}'C\hat{C}} \widetilde{J_2}= (c_1\hat{c}_2+c_2\hat{c}_1)I_{B\hat{B}}$.
    It is straightforward to check that $\{\widetilde{c}_1,\widetilde{c}_2,\widetilde{J_1},\widetilde{J_2}\}$
    is a feasible solution for $\rho_{ABC}\ox\sigma_{\hat{A}\hat{B}\hat{C}}$. For `$\geq$': For simplicity, we denote $\{K_1,M_1,N_1\}$ as a feasible solution for $\rho_{ABC}$ and $\{K_2,M_2,N_2\}$ a feasible solution for $\sigma_{\hat{A}\hat{B}\hat{C}}$. Then we let
    \begin{equation*}
        \begin{aligned}
            \widetilde{K} &= K_1\ox K_2,\\
            \widetilde{N} &= \frac{1}{2}\Big(\tr_A(K_1 \rho_{AB}^{T_B}) \ox (N_2 - M_2) + (N_1 - M_1) \ox\tr_A(K_2 \sigma_{\hat{A}\hat{B}}^{T_{\hat{B}}}) + N_1\ox N_2 + M_1\ox M_2\Big),\\
            \widetilde{M} &= \frac{1}{2}\Big(\tr_A(K_1 \rho_{AB}^{T_B}) \ox (M_2 - N_2) + (M_1 - N_1) \ox \tr_A(K_2 \sigma_{A\hat{B}}^{T_{\hat{B}}}) + N_1\ox M_2 + M_1\ox N_2\Big).
        \end{aligned}
    \end{equation*}
    It follows that
    \begin{equation*}
    \begin{aligned}
       &2 \left[\tr_{A\hat{A}}(K_1\ox K_2)(\rho_{AB} \ox \sigma_{\hat{A}\hat{B}})^{T_{B\hat{B}}} + \widetilde{N} \right]\\
       = & (\tr_A(K_1 \rho_{AB}^{T_B}) + N_1)\ox (\tr_{\hat{A}}(K_2 \sigma_{\hat{A}\hat{B}}^{T_{\hat{B}}}) + N_2) + (-\tr_A(K_1 \rho_{AB}^{T_B}) + M_1)\ox (-\tr_{\hat{A}}(K_2 \sigma_{\hat{A}\hat{B}}^{T_{\hat{B}}}) + M_2) \geq 0,\\
       &2 \left[- \tr_{A\hat{A}}(K_1\ox K_2)(\rho_{AB} \ox \sigma_{\hat{A}\hat{B}})^{T_{B\hat{B}}} + \widetilde{M} \right]\\
       = & (\tr_A(K_1 \rho_{AB}^{T_B}) + N_1)\ox (-\tr_A(K_2 \sigma_{\hat{A}\hat{B}}^{T_{\hat{B}}}) + M_2) + (\tr_A(K_2 \sigma_{\hat{A}\hat{B}}^{T_{\hat{B}}}) + N_2)\ox (-\tr_A(K_1 \rho_{AB}^{T_B}) + M_1)\geq 0,
   \end{aligned}
    \end{equation*}
    where we omit the labels of systems and the identity operators. Besides, we have 
    \begin{equation}
    \begin{aligned}
    &\tr \widetilde{M} = \frac{1}{2}(\tr N_1\tr N_2 + \tr M_1 \tr M_2) = 1,\\
    &\tr \widetilde{N} = \frac{1}{2}(\tr N_1\tr M_2 + \tr M_1 \tr N_2) = 1.
    \end{aligned}
    \end{equation}
    Hence, $\{\widetilde{K},\widetilde{M},\widetilde{N}\}$ is a feasible solution for $\rho_{ABC}\ox\sigma_{\hat{A}\hat{B}\hat{C}}$ which gives $2^{\nu(\rho_{ABC}\ox\sigma_{\hat{A}\hat{B}\hat{C}})} \geq 2^{\nu(\rho_{ABC})} \cdot 2^{\nu(\sigma_{\hat{A}\hat{B}\hat{C}})}$. In conclusion, we have proved $2^{\nu(\rho_{ABC}\ox\sigma_{\hat{A}\hat{B}\hat{C}})} = 2^{\nu(\rho_{ABC})} \cdot 2^{\nu(\sigma_{\hat{A}\hat{B}\hat{C}})}$ which yields 
    \begin{equation}
        \nu(\rho_{ABC}\ox\sigma_{\hat{A}\hat{B}\hat{C}}) = \nu(\rho_{ABC}) + \nu(\sigma_{\hat{A}\hat{B}\hat{C}}).
    \end{equation}
\end{proof}

The additivity of the virtual non-Markovianity with respect to the tensor product of quantum states implies that for parallel corrupted states, a global recovering protocol has no advantage over a local recovering protocol, i.e., recovering each state individually.

\begin{figure*}[t]
    \centering
    \includegraphics[width=.9\linewidth]{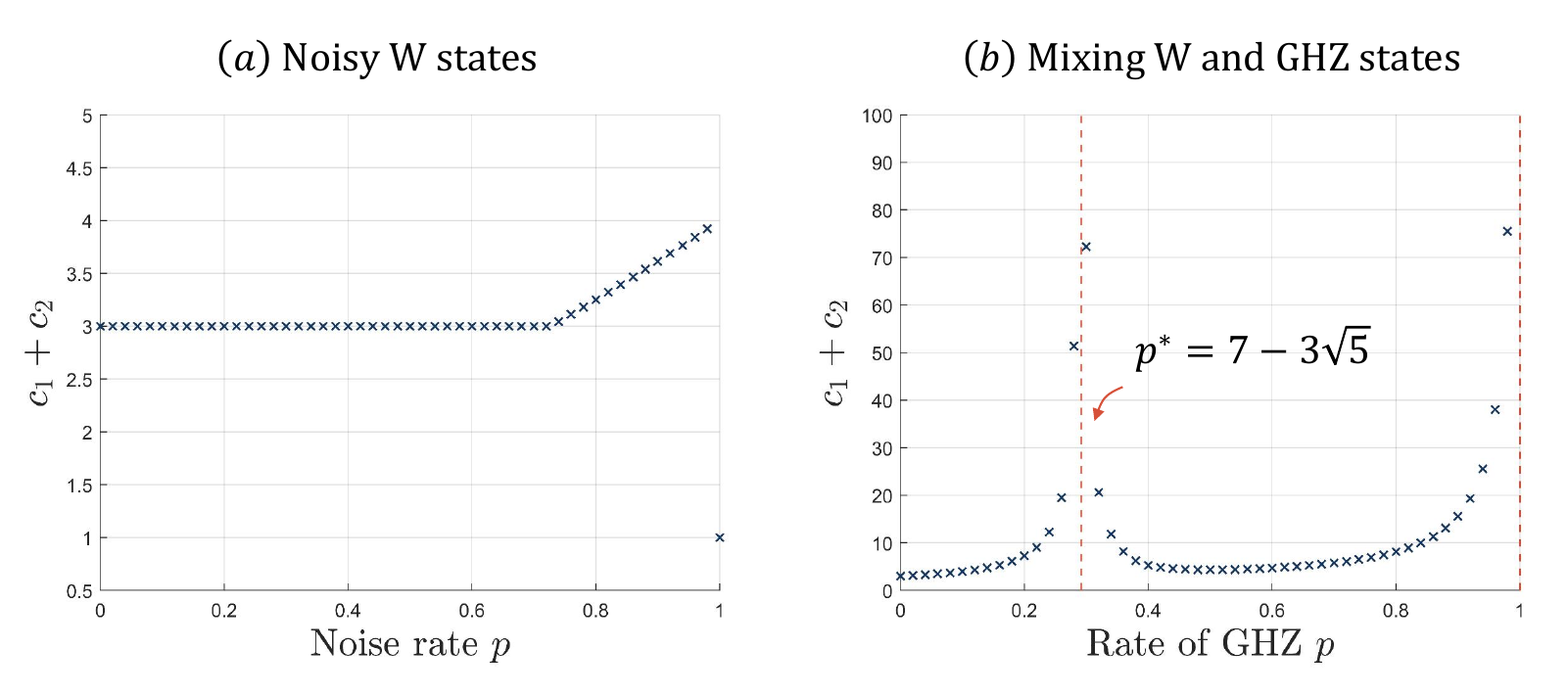}
    \caption{The virtual non-Markovianity of W state mixing with other quantum states. Panel (a) illustrates the case of virtual recovering a 3-qubit W state affected by a depolarizing channel of noise rate $p$; Panel (b) illustrates the case of virtual recovering $GW(p)$, a W state mixed with a GHZ state. At $p^* = 7-3\sqrt{5}$, the mixed state is not a VQMC.}
    \label{fig:mixing_w_and_other_states}
\end{figure*}

We investigate the virtual non-Markovianity of different types of tripartite quantum states. Firstly, consider a W state under depolarizing noise as defined in Eq.~\eqref{Eq:depo_W_GHZ}.
We present the virtual non-Markovianity of $W(p)$ in Fig.~\ref{fig:mixing_w_and_other_states} with $p\in[0,1]$. We observe that the virtual non-Markovianity is $\log(c_1 + c_2) = \log 3$ when $0\leq p\lesssim 0.725$. There is a jump discontinuity at $p=1$, indicating that even with mixing the maximally mixed state with an extremely small amount of W state, the virtual non-Markovianity increases a lot. Secondly, consider the convex combination of a three-qubit W state and a GHZ state as defined in Example~\ref{exm:ghz_mix_w}. The virtual non-Markovianity of $GW(p)$ is depicted in Fig.~\ref{fig:mixing_w_and_other_states} with $p\in[0,1]$. As stated in Example~\ref{exm:ghz_mix_w}, we observe that the virtual non-Markovianity is infinity when $p^*=7-3 \sqrt{5}$.

\section{Approximate virtual quantum Markov chain}\label{sec:approx vqmc}
In the same spirit as the approximate quantum Markov chain, we study whether the properties of virtual quantum Markov chains are robust in this section. Note that the information we want to recover for a virtual quantum Markov chain is its measurement statistics with respect to any observable, and the recovery map is not completely positive. We herein introduce the \textit{$\varepsilon$-approximate virtual quantum Markov chain}.

\begin{definition}[$\varepsilon$-approximate virtual quantum Markov chain]
    A tripartite quantum state $\rho_{ABC} \in \cD(A\ox B\ox C)$ is called an $\varepsilon$-approximate virtual quantum Markov chain in order $A\leftrightarrow B\leftrightarrow C$ if 
    \begin{equation}\label{def:approx_vqmc}
        \min_{\mathscr{M}\in\HPTP(B,B\ox C)} \left\|\mathscr{M}_{B\rightarrow BC}\circ \tr_C(\rho_{ABC}) - \rho_{ABC}\right\|_1 = \varepsilon,
    \end{equation}
    where $\mathscr{M}_{B\rightarrow BC}$ ranges over all HPTP maps.
\end{definition}

Denote the optimal map in Eq.~\eqref{def:approx_vqmc} as $\widetilde{\mathscr{M}}_{B\rightarrow BC}$ and $\widehat{\rho}_{ABC} = \mathscr{M}_{B\rightarrow BC}\circ \tr_C(\rho_{ABC})$. Using a quasiprobability decomposition implementation for $\widetilde{\mathscr{M}}_{B\rightarrow BC}$, we can estimate the value of $\tr(O\rho_{ABC})$ approximately for any possible observable $O$. Specifically, for any given observable $O$, we have
\begin{equation}\label{Eq:upperbound_approx_vqmc}
|\tr(O\widehat{\rho}_{ABC}) - \tr(O\rho_{ABC})| \leq \|O\|_{\infty}\cdot \left\|\widehat{\rho}_{ABC} - \rho_{ABC}\right\|_1 = \|O\|_{\infty}\cdot\varepsilon,
\end{equation}
where the inequality is followed by Hölder's inequality.
The $\varepsilon$ actually corresponds to the approximate virtual recoverability of $\rho_{ABC}$. For any given tripartite quantum state $\rho_{ABC}$, its approximate virtual recoverability can be evaluated by the following SDPs.

\begin{equation}\label{sdp:tr_norm_pri_dual}
\begin{aligned}
&\underline{\textbf{Primal Program}}\\
\min &\;\; \tr S_{ABC} \\
{\rm s.t.}
&\;\; \tr_B(\rho_{AB}^{T_B}\ox I_{B'C})(I_A\ox J_{BB'C}) = \sigma_{AB'C}, \\
&\;\; S_{ABC} \geq \rho_{ABC} - \sigma_{ABC},~S_{ABC}\geq 0,\\
&\;\; \tr_{B'C} J_{BB'C} = I_B, \\
\end{aligned}
\quad
\begin{aligned}
&\underline{\textbf{Dual Program}}\\
\max & \; \tr P_{B} + \tr [R_{ABC}\rho_{ABC}] \\
{\rm s.t.}
&\;\; \tr_A (R_{AB'C}\rho_{AB}^{T_B}) = \tau_{BB'C},\\
&\;\; P_B \ox I_{B'C} +\tau_{BB'C}  = 0,\\
&\;\; R_{ABC} \leq I_{ABC}.
\end{aligned}
\end{equation}

\begin{figure}[t]
    \centering
    \includegraphics[width=.5\linewidth]{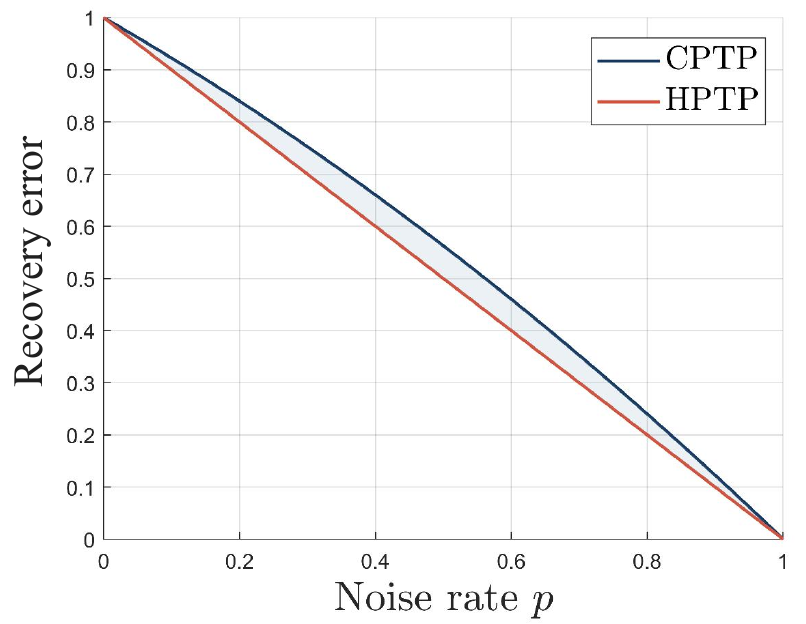}
    \caption{Comparison of $\varepsilon$ under CPTP and HPTP recovery maps. A 3-qubit GHZ state is affected by depolarizing channels with varying noise rates $p$. The dark blue and red curves illustrate the minimum $\varepsilon$ values assisted by CPTP and HPTP, respectively.}
    \label{fig:hptp_vs_cptp_depghz}
\end{figure}

\begin{figure}
    \centering
    \includegraphics[width=.5\linewidth]{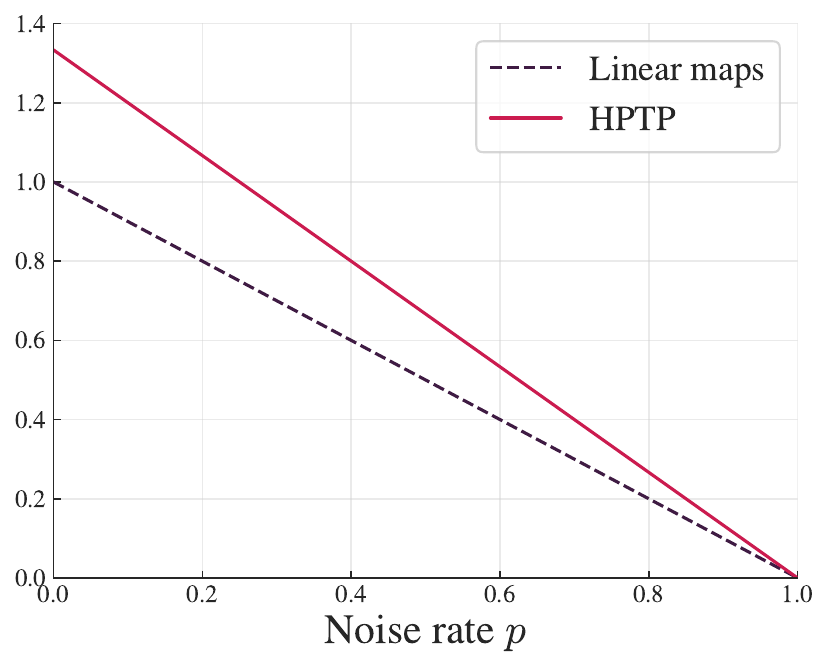}
    \caption{Comparison of reconstruction errors using different map types. The dashed line shows the minimum trace distance $\varepsilon_{\rm Lin}(\rho_p)$ achievable by general linear maps, while the red line shows the minimum trace distance $\varepsilon_{\rm HPTP}(\rho_p)$ achievable by HPTP maps, both plotted against the mixing parameter $p$. As $p$ increases, the state becomes more mixed, and the difference between the two approaches becomes more pronounced.}
    \label{fig:LvsHPTP}
\end{figure}

The derivation of the dual SDP is in Appendix~\ref{appendix:sdp_approximate-virtual-recoverability}. We consider a tripartite state $G(p)$ as defined in Eq.~\eqref{Eq:depo_W_GHZ} and utilize SDPs in Eq.~\eqref{sdp:tr_norm_pri_dual} to determine the approximate virtual recoverability of $G(p)$ with $p\in [0,1]$. Note that we can also utilize the primal SDP in Eq.~\eqref{sdp:tr_norm_pri_dual} to calculate the minimum $\varepsilon$ when the recovery map $\mathscr{M}_{B\rightarrow BC}$ allowed is restricted to CPTP by simply adding a positivity constraint for $J_{BB'C}$ ($J_{BB'C}\geq 0$). We show the minimum $\varepsilon$ for $G(p)$ with different noise parameters $p$ in Fig.~\ref{fig:hptp_vs_cptp_depghz} where the red line corresponds to the value of Eq.~\eqref{def:approx_vqmc} and the blue line corresponds to the case where $\mathscr{M}_{B\rightarrow BC}$ ranges over all quantum channels. We observe that HPTP maps can indeed offer improvement, considering a general upper bound $\varepsilon\cdot \|O\|_\infty$ in Eq.~\eqref{Eq:upperbound_approx_vqmc} for the expectation value with respect to any observable.

\begin{remark}
We would like to remark on the difference between HPTP maps and linear maps in terms of the reconstruction errors. Consider a three-qutrit noisy GHZ state
\begin{equation}
    \rho_p = (1-p)\ketbra{\psi}{\psi} + p\frac{\mathbb{I}}{27},
\end{equation}
where $\ket{\psi} = \frac{1}{\sqrt 3}(\ket{000}+\ket{111} + \ket{222})$. We analyze the minimum trace distance achievable by HPTP maps:
\begin{equation}
    \varepsilon_{\rm HPTP}(\rho_p) = \min_{\mathscr{M}\in\mathrm{HPTP}(B,B\ox C)} \left\|\mathscr{M}_{B\rightarrow BC}\circ \tr_C(\rho_p) - \rho_p\right\|_1,
\end{equation}
and the minimum trace distance achievable by general linear maps:
\begin{equation}
    \varepsilon_{\rm Lin}(\rho_p) = \min_{\cN\in\cL(B,B\ox C)} \left\|\cN_{B\rightarrow BC}\circ \tr_C(\rho_p) - \rho_p\right\|_1.
\end{equation}
We can see the gap in the reconstruction errors between the two types of maps in Fig.~\ref{fig:LvsHPTP}. Thus, it is meaningful and fundamentally important to explore the relationship of different classes of maps in the task of reconstruction or state recovery, especially for certain classes such as HPTP, which is physically relevant~\cite{Temme2017,Jiang2020,Piveteau2021,Yao2023,Zhao2023,Zhao2022,Zhu_2024,Parzygnat_2024}. The codes for numerical experiments are available at~\cite{coderepo}.
\end{remark}

\section{Discussions}
To deepen our understanding of how the lost information of a quantum system can be recovered from a correlated subsystem, we have introduced the \textit{virtual quantum Markov chains}, which allow for the recovery of global shadow information from subsystems via quantum operations and post-processing. An algebraic characterization of virtual quantum Markov chains is provided, and the optimal sampling overhead for the virtual quantum recovery can be efficiently computed via SDP, based on which we present a quantifier for the non-Markovianity of quantum states named the virtual non-Markovianity. Furthermore, we introduce the approximate virtual quantum Markov chains, where shadow information with respect to any observable can be recovered approximately. It admits an advantage over conventional methods using quantum channels.
Our results shed light on the quantification of non-Markovianity (the memory effects exhibited by quantum systems) and inspire potential applications to distributed quantum computing, quantum error mitigation, and entanglement wedge reconstruction~\cite{penington2020entanglement}.

Further investigation into the properties and applications of approximate VQMC presents a compelling avenue of research, e.g., an entropic characterization~\cite{Petz1986,Fawzi2014}. Additionally, an intriguing direction worth pursuing involves the development of protocols akin to the Petz recovery map and universal recovery maps~\cite{Sutter2016c}, but specifically tailored for VQMC. It is worth noting that the sampling complexity of reconstructing a quantum state through state tomography can be reduced if the state is known to be a QMC~\cite{gao2022}. Considering that a QMC is a special subclass of VQMC, it would also be intriguing to explore whether other classes of VQMC can be beneficial in learning density matrices and to determine the minimal prior information required about VQMCs in learning tasks. These paths of exploration could advance the understanding of the limits of recoverability in quantum information theory.

\section*{Acknowledgment}
We thank the anonymous referee for pointing out an error in our original manuscript about classical VQMC, providing a counterexample demonstrating that a classical state does not necessarily imply a VQMC, and informing the reference~\cite{Holz_pfel_2018}. This work was partially supported by the National Key R\&D Program of China (Grant No. 2024YFE0102500), the National Natural Science Foundation of China (Grant. No.~12447107), the Guangdong Provincial Quantum Science Strategic Initiative (Grant No.~GDZX2403008, GDZX2403001), the Guangdong Provincial Key Lab of Integrated Communication, Sensing and Computation for Ubiquitous Internet of Things (Grant No. 2023B1212010007), the Quantum Science Center of Guangdong-Hong Kong-Macao Greater Bay Area, and the Education Bureau of Guangzhou Municipality.


\appendix
\setcounter{subsection}{0}
\setcounter{table}{0}
\setcounter{figure}{0}


\renewcommand{\theequation}{S\arabic{equation}}
\renewcommand{\thesubsection}{\normalsize{Supplementary Note \arabic{subsection}}}
\renewcommand{\theproposition}{S\arabic{proposition}}
\renewcommand{\thedefinition}{S\arabic{definition}}
\renewcommand{\thefigure}{S\arabic{figure}}
\setcounter{equation}{0}
\setcounter{table}{0}
\setcounter{section}{0}
\setcounter{proposition}{0}
\setcounter{definition}{0}
\setcounter{figure}{0}

\section{Remarks on Theorem~\ref{thm:main_necc_suff}}

In this section, we remark how Theorem~\ref{thm:main_necc_suff} can be applied to see that both classical Markov chains and quantum Markov chains are special cases of virtual quantum Markov chains (VQMCs).

\begin{proposition}\label{lem:qmc_to_VQMC}
If a tripartite quantum state $\rho_{ABC}$ is a quantum Markov chain in order $A\leftrightarrow B\leftrightarrow C$, then it is a virtual quantum Markov chain.
\end{proposition}
\begin{proof}
Recall that $\rho_{ABC}$ is a quantum Markov chain if and only if system $B$ can be decomposed into a direct sum tensor product
\begin{equation}
    \cH_B = \underset{t}{\bigoplus}  \cH^{(t)}_{b^L} \ox \cH^{(t)}_{b^R}, ~ \text{s.t.} \: \rho_{ABC} = \underset{t}{\bigoplus} q_t \rho^{(t)}_{Ab^L} \ox \rho^{(t)}_{b^RC},
\end{equation} 
with states $\rho_{Ab^L_j}$ on $\mathcal{H}_A \ox \mathcal{H}_{b^L_j}$ and $\rho_{b^R_jC}$ on $ \cH_{b^R_j} \ox \mathcal{H}_C $ and a probability distribution $\{q_t\}$.
If $\rho_{ABC} = \underset{t}{\bigoplus} q_t \rho^{(t)}_{Ab^L} \ox \rho^{(t)}_{b^R C}$, we have
\begin{equation}
\begin{aligned}
    Q_{BC}^{(ij)} &= \bigoplus_t q_t (\bra{i}_A\ox I_{b^L})\rho^{(t)}_{Ab^L}(\ket{j}_A\ox I_{b^R}) \ox \rho^{(t)}_{b^R C},\\
    Q_{B}^{(ij)} &= \bigoplus_t q_t (\bra{i}_A\ox I_{b^L})\rho^{(t)}_{Ab^L}(\ket{j}_A\ox I_{b^R}) \ox \rho^{(t)}_{b^R}.
\end{aligned}
\end{equation}
Thus, for any $\mathbf{c}\in \ker\RecB$ such that 
\begin{equation}
\begin{aligned}
    \mathbf{0} &= \sum_{ij} c_{ij} Q^{(ij)}_{B}\\
    &= \sum_{ij} c_{ij} \bigoplus_t q_t (\bra{i}_A\ox I_{b^L})\rho^{(t)}_{Ab^L}(\ket{j}_A\ox I_{b^R}) \ox \rho^{(t)}_{b^R}\\
    &= \bigoplus_t q_t \left(\sum_{ij} c_{ij} (\bra{i}_A\ox I_{b^L})\rho^{(t)}_{Ab^L}(\ket{j}_A\ox I_{b^R}) \right) \ox \rho^{(t)}_{b^R},
\end{aligned}
\end{equation}
it follows that 
\begin{equation}
    \sum_{ij} c_{ij} (\bra{i}_A\ox I_{b^L})\rho^{(t)}_{Ab^L}(\ket{j}_A\ox I_{b^R}) = 0.
\end{equation}
Hence we have
\begin{equation}
\begin{aligned}
    \sum_{ij} c_{ij} Q_{BC}^{(ij)}
    =& \sum_{ij} c_{ij} \bigoplus_t q_t (\bra{i}_A\ox I_{b^L})\rho^{(t)}_{Ab^L}(\ket{j}_A\ox I_{b^R})\ox \rho^{(t)}_{b^RC} \\
    =& \bigoplus_t q_t \left(\sum_{ij} c_{ij} (\bra{i}_A\ox I_{b^L})\rho^{(t)}_{Ab^L}(\ket{j}_A\ox I_{b^R}) \right) \ox \rho^{(t)}_{b^RC}\\
    =& 0,
\end{aligned}
\end{equation}
which yields $\mathbf{c}\in\ker\RecBC$. Applying Theorem~\ref{thm:main_necc_suff}, we conclude.
\end{proof}

\begin{proposition}\label{prop:cmc_to_VQMC}
If a tripartite classical state $\rho_{ABC}$ is a Markov chain in order $A\leftrightarrow B\leftrightarrow C$, then it is a virtual quantum Markov chain.
\end{proposition}
\begin{proof}
For a classical state $\rho_{ABC} = \sum_{ijk} p_{ijk} \ketbra{i}{i} \ox \ketbra{j}{j} \ox \ketbra{k}{k}$, we have
\begin{equation}
    Q_{BC}^{(ii)} = \sum_{jk} p_{ijk} \ketbra{j}{j} \ox \ketbra{k}{k}, ~ Q_{B}^{(ii)} = \sum_{j} p_{ij} \ketbra{j}{j},
\end{equation}
where $p_{ij} = \sum_{k}p_{ijk}$. For any $\mathbf{c}\in \ker\RecB$, we have 
\begin{equation}
    \sum_{i} c_i p_{ijk} = \frac{p_{jk}}{p_j} \sum_{i} c_i p_{ij} = 0, \, \forall j,
\end{equation}
where the first equality is by the fact that $p_{ijk} = p_{ij}\cdot p_{jk}/p_{j}$ for a classical Markov chain, and the second equality is because $\mathbf{c}\in \ker\RecB$. Then it follows that $\mathbf{c}\in\ker\RecBC$. By Theorem~\ref{thm:main_necc_suff}, we have $\rho_{ABC}$ is a virtual quantum Markov chain.
\end{proof}

\section{SDP for virtual non-Markovianity}\label{appendix:sdp_sampling}
In this section, we provide a derivation of the dual SDP for calculating the virtual non-Markovianity. Recall that the primal SDP for the virtual non-Markovianity of $\rho_{ABC}$ can be written as follows, where $J_{BB'C}$ is the recovery map that can be written as a linear combination of two CPTP maps $J_1$ and $J_2$.
\begin{subequations}
\begin{align}\label{sdp:primal_cost}
2^{\nu(\rho_{ABC})}&\\
= \min & \; c_1 + c_2\\
 {\rm s.t.} & \;\; J_{1}\geq 0, J_{2}\geq 0, \\
    & \;\; J_{BB'C} = J_1 - J_2,\\
    &\;\; \tr_{B'C} J_{1} = c_1 I_B,\tr_{B'C} J_{2} = c_2 I_B,\\
    & \;\; \tr_B(\rho_{AB}^{T_B}\ox I_{B'C})(I_A\ox J_{BB'C}) = \rho_{ABC}.
\end{align}
\end{subequations}
The Lagrange function of the primal problem is
\begin{equation}
\begin{aligned}
    &L(M, N, K, c_1,c_2,J_1,J_2)\\ 
    =\;& c_1 + c_2 + \langle M, \tr_{B'C}J_1 - c_1 I_B\rangle + \langle N, \tr_{B'C}J_2 - c_2 I_B\rangle\\
    & + \langle K, \rho_{ABC} - \tr_B(\rho_{AB}^{T_B}\ox I_{B'C})(I_A\ox J_{BB'C})\rangle\\
    =\;& \tr(K_{AB'C}\rho_{AB'C}) + c_1(1-\tr M) + c_2 (1-\tr N)\\
    & + \langle J_1, M_B\ox I_{B'C}-\tr_A[(K_{AB'C}\ox I_B)(\rho_{AB}^{T_B}\ox I_{B'C})]\rangle\\
    & + \langle J_2, N_B\ox I_{B'C}+\tr_A[(K_{AB'C}\ox I_B)(\rho_{AB}^{T_B}\ox I_{B'C})]\rangle,
\end{aligned}
\end{equation}
where $M_{B}, N_{B}, K_{AB'C}$ are Lagrange multipliers. The corresponding Lagrange dual function is 
\begin{equation}
    g(M,N,K) = \inf_{J_1,J_2\geq 0, c_1,c_2\geq 0} L(M, N, K, c_1,c_2,J_1,J_2).
\end{equation}
Since $J_1 \geq 0, J_2 \geq 0$, it must hold that $\tr M \leq 1, \tr N \leq 1$ and 
\begin{equation*}
\begin{aligned}
    M_B\ox I_{B'C} - \tr_A[(K_{AB'C}\ox I_B)(\rho_{AB}^{T_B}\ox I_{B'C})] &\geq 0,\\
    N_B\ox I_{B'C} + \tr_A[(K_{AB'C}\ox I_B)(\rho_{AB}^{T_B}\ox I_{B'C})] &\geq 0.
\end{aligned}
\end{equation*}
Thus, the dual SDP is
\begin{equation}
\begin{aligned}\label{sdp:state_dual}
&\max_{M_{B},N_{B},K_{AB'C}}  \tr[K_{AB'C}\rho_{ABC}]\\
 {\rm s.t.} &\; \tr M_{B} \leq 1,\tr N_{B} \leq 1, \\
&\; M_B\ox I_{B'C}- \tr_A[(K_{AB'C}\ox I_B)(\rho_{AB}^{T_B}\ox I_{B'C})] \geq 0,\\
&\; N_B\ox I_{B'C}+ \tr_A[(K_{AB'C}\ox I_B)(\rho_{AB}^{T_B}\ox I_{B'C})] \geq 0.
\end{aligned}
\end{equation}

\section{SDP for approximate virtual recoverability}\label{appendix:sdp_approximate-virtual-recoverability}
In this section, we provide a detailed SDP derivation for the approximate virtual recoverability. The primal SDP for approximate virtual recoverability can be written as:
\begin{equation}
    \begin{aligned}
    \min_{S_{ABC} \geq 0, J_{BB'C}}  & \tr S_{ABC} \\
        {\rm s.t.}   &\;\; S_{ABC} \geq \rho_{ABC} - \sigma_{ABC}, \\
        &\;\; \tr_B(\rho_{AB}^{T_B}\ox I_{B'C})(I_A\ox J_{BB'C}) = \sigma_{AB'C},\\
        &\;\; \tr_{B'C} J_{BB'C} = I_B. \\
    \end{aligned}
\end{equation}
The Lagrange function of the primal problem is
\begin{equation*}
\begin{aligned}
    &L(P, R,S, J) \\
    =& \tr S_{ABC} + \tr [P_B(I_B - \tr_{B'C} J_{BB'C})]\\ 
    &\quad + \big\langle R_{ABC}, \rho_{ABC} - \tr_B (\rho_{AB}^{T_B} \otimes I_{B'C})(I_A \otimes J_{BB'C}) - S_{ABC}]\big\rangle\\ 
    =& \tr P_{B} + \tr [R_{ABC}\rho_{ABC}] + \tr[S_{ABC}(I_{ABC} - R_{ABC})]\\
    &\quad - \big\langle J_{BB'C}, P_B \otimes I_{B'C} + \tr_A [(R_{AB'C} \otimes I_B)(\rho_{AB}^{T_B} \otimes I_{B'C})] \big\rangle,     
\end{aligned}
\end{equation*}
where $P_B, R_{ABC}$ are Lagrange multipliers. $S_{ABC} \geq 0$ and $J_{BB'C} \in \mathcal{L}^\dagger(\mathcal{H})$ respectively require
\begin{equation}
\begin{aligned}
    &I_{ABC} - R_{ABC} \geq 0, \\
    &P_B \otimes I_{B'C} + \tr_A [R_{AB'C} \otimes I_B(\rho_{AB}^{T_B} \otimes I_{B'C})] = 0.
\end{aligned}
\end{equation}
Thus, the dual SDP is
\begin{equation}
\begin{aligned}
\max_{R_{ABC} \geq 0, P_{B}} & \; \tr P_{B} + \tr [R_{ABC}\rho_{ABC}] \\
    {\rm s.t.}   &\;\; I_{ABC} \geq R_{ABC}, \\
    &\;\; P_B \otimes I_{B'C} + \tr_A [(R_{AB'C} \otimes I_B)(\rho_{AB}^{T_B} \otimes I_{B'C})] = 0.        
\end{aligned}
\end{equation}

\begin{figure}[htbp]
    \centering
    \includegraphics[width=.5\linewidth]{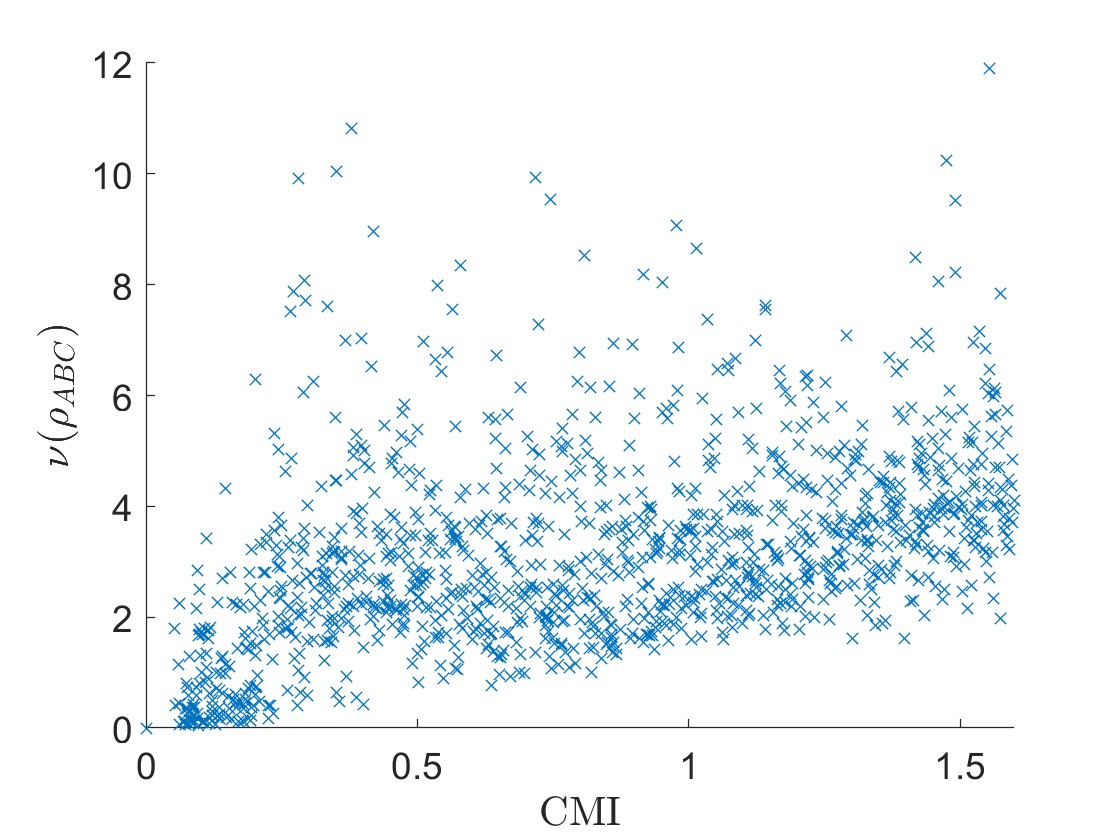}
    \caption{The virtual non-Markovianity vs conditional mutual information.}
    \label{fig:cost_vs_CMI_random}
\end{figure}

\section{Relation between the virtual non-Markovianity and the conditional mutual information.}\label{appendix:VNM_vs_CMI}
In this section, we show some numerical results on the potential connection between the virtual non-Markovianity and the conditional mutual information of a tripartite quantum state.

\paragraph{Random states}
We randomly sample 3-qubit quantum states to obtain 1500 ones such that their conditional mutual information is nearly uniformly distributed between 0 and 1.5. Specifically, we ensure that there are 50 randomly sampled quantum states within each 0.05 interval of the conditional mutual information. This process is done by rejection sampling, e.g., once there are 50 samples in an interval, additional ones are rejected. Then we calculate the virtual non-Markovianity of each state. The result is depicted in Fig.~\ref{fig:cost_vs_CMI_random}. As the conditional mutual information increases, there is a clear upward trend in the virtual non-Markovianity. There may exist some inequality relation between these two quantities, such as a function of the conditional mutual information would give a lower bound on virtual non-Markovianity.

\begin{figure}[htbp]
    \centering
    \includegraphics[width=.5\linewidth]{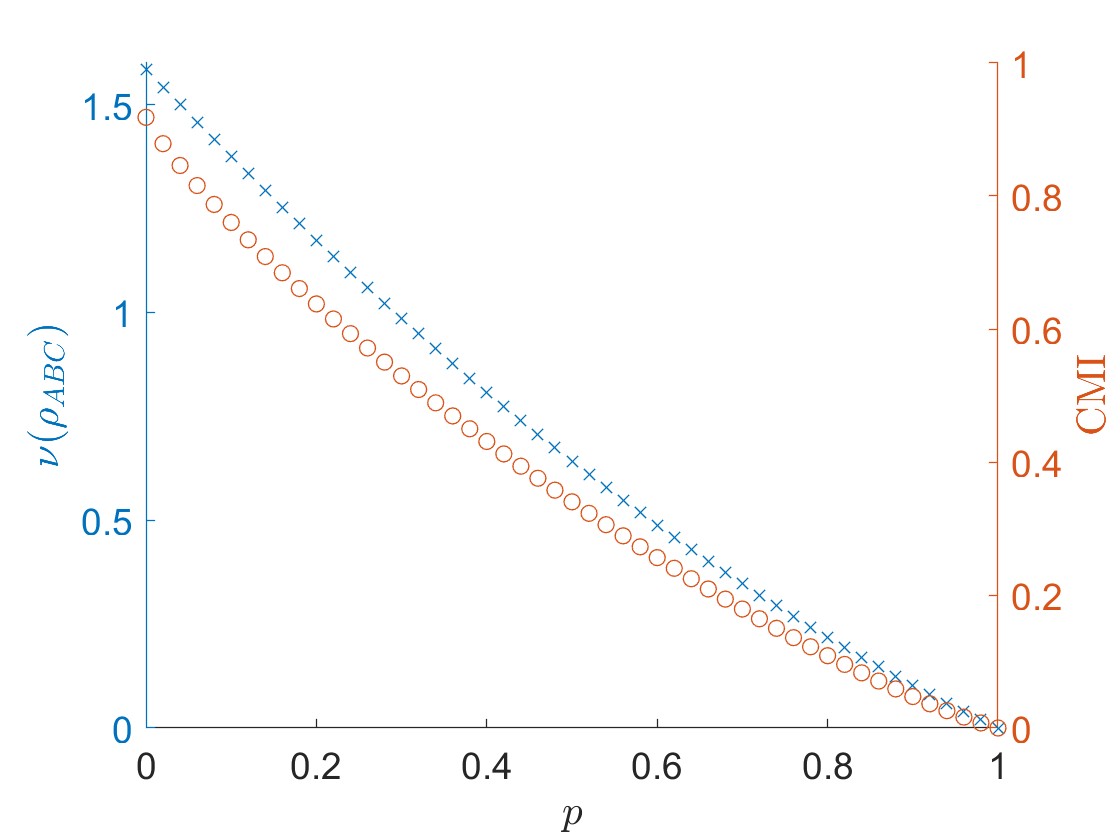}
    \caption{The virtual non-Markovianity and CMI of W state mixing with $\ket{\Psi^+}\bra{\Psi^+} \otimes I/2$ where $\ket{\Psi^+} = (\ket{01} + \ket{10})/\sqrt{2}$. }
    \label{fig:MixW_vs_CMI_v}
\end{figure}

\paragraph{Specific state}
We also consider another explicit example. For a tripartite state $\rho_{ABC} = (1-p) \ketbra{W}{W} + p \ketbra{\Psi^+}{\Psi^+}\ox I_2$ where $\ket{\Psi^+} = (\ket{01} + \ket{10})/\sqrt{2}$, we calculate its virtual non-Markovianity and conditional mutual information when $p$ varies. The result is shown in Fig.~\ref{fig:MixW_vs_CMI_v}. We can observe that when both quantities decrease as $p$ increases and there exists a potential relationship between these quantities.

\bibliographystyle{ieeetr}
\bibliography{main}

\end{document}